\documentclass[12pt]{article}

\usepackage{indentfirst,natbib,graphics,graphicx,epsfig,amssymb,color}

\begin{document}

\newtheorem{teo}{Theorem}
\newtheorem{deh}{Definition}
\newtheorem{coc}{Corollary}
\newtheorem{pop}{Proposition}
\newtheorem{rmk}{Remark}
\newtheorem{lem}{Lemma}
\newtheorem{coco}{Conjecture}
\newcommand{\betac}{\beta_{t_1, t_2; j_1, j_2}^{(f)}}
\newcommand{\gammaw}{\gamma_{t_1, t_2; j}^{(f)}}
\newcommand{\Corr}{{\rm Corr}}
\newcommand{\Cov}{{\rm Cov}}
\newcommand{\E}{{\rm E}}
\newcommand{\pr}{{\rm Pr}}
\newcommand{\V}{{\rm Var}}
\newcommand{\bq}{\begin{quote}}
\newcommand{\eq}{\end{quote}}
\newcommand{\beq}{\begin{equation}}
\newcommand{\eeq}{\end{equation}}
\newcommand{\beqn}{\begin{eqnarray}}
\newcommand{\eeqn}{\end{eqnarray}}
\newcommand{\Uni}{{\rm Uniform}}
\newcommand{\CN}{{\cal N}}
\newcommand{\CS}{{\cal S}}
\newcommand{\CH}{{\cal H}}
\newcommand{\CX}{{\cal X}}
\newcommand{\CA}{{\cal A}}
\newcommand{\CB}{{\cal B}}
\newcommand{\M}{{\cal M}}

\newcommand{\CU}{{\cal U}^k}
\newcommand{\CUTW}{{\cal U}^2}
\newcommand{\dd}{{\cal D}}
\newcommand{\CV}{{\cal V}^k}
\newcommand{\CK}{{\cal K}}
\newcommand{\rcov}{\mbox{Cov}}
\newcommand{\rcorr}{\mbox{Corr}}
\newcommand{\rvar}{\mbox{Var}}
\newcommand{\re}{\mbox{E}}
\newcommand{\ru}{\mbox{Uniform}}
\newcommand{\RR}{{\bf R}}
\newcommand{\II}{{\bf I}}
\newcommand{\BX}{{\bf X}}
\newcommand{\BY}{{\bf Y}}
\newcommand{\rep}{14}

\def\supp{{\rm supp}}
\def\eqref#1{(\ref{#1})}

\title{Additive Models for Conditional Bivariate Copulas: A Bayesian Approach}

\author{Avideh Sabeti$^\ddagger$, Mian Wei$^\ddagger$ and Radu ~V.~Craiu$^\ddagger$\\
$^\ddagger$ Department of Statistics, University of Toronto, Canada}

\maketitle

\begin{abstract}
Conditional copulas are flexible statistical tools that couple joint  conditional  and marginal conditional distributions. In a linear regression setting with more than one covariate and two dependent outcomes,  we propose the use of additive models for conditional bivariate copula models and discuss computation and model selection tools for performing Bayesian inference. The method is illustrated using simulations and a real example.

\end{abstract}

\noindent Keywords: {\it Additive models, Bayesian inference, Cross-validated marginal likelihood, Conditional copulas, Cubic splines, Markov chain Monte Carlo.}

\section{Introduction}

Starting with the  seminal paper of \cite{Sklar:1959}, copulas have developed into an important tool used for modelling dependence in statistical models.  If $Y_{1}, Y_{2},\ldots, Y_k$ are   continuous random variables with joint distribution function $H$ and marginal distributions $F_{1},F_{2},\ldots,F_k$, the unique copula $C:[0,1]^{k} \to[0,1]$  ``couples" the  joint and the marginal distributions via 
 $ H(y_{1},\ldots,y_{k}) =  C \{F_{1}(y_{1}), \ldots F_{n}(y_{k}) \}$, for all $(y_{1},\ldots,y_{k}) \in\mathbb{R}^{k}$.  
Therefore, in order to define $H$,  we need the marginals $F_i$ and the copula $C$. This can be  convenient in situations in which one has a good grasp on  the marginal distributions. 

As a natural extension, conditional copulas  couple  joint  conditional and  marginal conditional distributions \citep{lamb-van,Patton:2006}. Specifically, if $X\in \RR^p$ is a covariate vector, then
\beq 
H_{X}(y_{1},\ldots,y_{k} \mid X)= C\{F_{1 \mid X}(y_{1} \mid X), \ldots,  F_{n \mid X}(y_{k} \mid X)  \mid X\} , \quad
\mbox{for all} \; (y_{1},\ldots,y_{k})  \in \RR^{k}. 
\label{cc:def}
\eeq

Conditional copulas models play an essential part in modelling high dimensional data. For instance, consider $X=(X_1,\ldots,X_4) \in \RR^4$. Using a similar  decomposition to the one used by  \cite{agn} (equation (3) at page 75) we  can show that its four-dimensional continuous density $f(x):=f(x_1,x_2,x_3,x_4)$ can be decomposed as 
\beqn
f(x_1,x_2,x_3,x_4) &=& f_1(x_1) f_2(x_2) f_3(x_3) f_4(x_4) \nonumber \\  
&\times& c_{12}\{F_1(x_1),F_2(x_2)\}  c_{23}\{F_2(x_2),F_3(x_3)\}  c_{14}\{F_1(x_1),F_4(x_4)\} \nonumber \\
&\times& c_{13\mid 2}\{F_{1 \mid 2}(x_1|x_2),F_{3 \mid 2}(x_3|x_2)\} c_{24\mid 1}\{F_{2 \mid 1}(x_2|x_1),F_{4 \mid 1}(x_4|x_1)\} \nonumber \\
&\times& c_{43\mid 12} \{F_{4\mid 12} (x_4\mid x_1,x_2), F_{3\mid 12}(x_3 \mid x_1,x_2)\},
\label{vine}
\eeqn 
where, if $\CA,\CB \subset \{1,2,3,4\}$ are set of indices and we have used the following notations: $f_{\CA}$, $F_{\CA}$ are, respectively, the joint density and distribution function of 
$\{X_j : j \in \CA\}$; $f_{\CA \mid \CB }$, $F_{\CA \mid \CB}$ are the   conditional density and distribution functions of $\{X_j : j \in \CA\}$ given $\{X_h : h \in \CB \}$; $c_{\CA}$  and  $c_{\CA\mid \CB}$ denote, respectively,  the copula density for $\{X_j : j \in \CA\}$ and the conditional copula density of $\{X_j : j \in \CA\}$ given $\{X_h : h \in \CB \}$. Not  surprisingly, increasing  the dimension of $X$ will result in a decomposition like (\ref{vine}) where we need to  condition on more than two random variables. \cite{agn} have shown that when replacing the conditional copulas with unconditional ones in (\ref{vine}), we are likely to incur inferential losses in terms of both bias and efficiency. 

The conditional copula can also be a useful modelling  tool  in regression settings in which we  observe outcomes $Y_1,\ldots, Y_k$ along with covariate vector $X \in \RR^p$ and of interest is not only the effect of the covariate on each response, but also the effect of $X$ on the dependence structure between the responses. Throughout the paper we consider parametric copula families in which  the function $C$ assumes a parametric form indexed by a copula parameter $\theta$.  In many applications one can reasonably assume  that $\theta$ will vary with $X$. However, it is generally difficult to guess the functional relationship between $\theta$ and the covariate vector $X$ so its estimation requires flexible models that can capture a wide variety of patterns.  This naturally leads to the use of semiparametric \citep{acy, cra-sabeti} and nonparametric inferential tools \citep{Omelka:2009ov,vog,abegaz}. As the dimension $p$ of the covariate vector $X$ increases, the volume of data required to keep the error within reasonable bounds increases very quickly \citep{abegaz}.  However, the generic examples discussed above motivate our search for practical inferential procedures for conditional copula models when $p>1$. The paper is developed situations in which the parameter $\theta$ is a scalar and  there are two (i.e. $k=2$) continuous outcomes of interest, $Y_1$ and $Y_2$, that are  marginally linked to the vector of covariates via linear regression models.

We propose here the use of additive models for studying the functional dependence between the covariate vector and the copula parameter. In this paper we will improve on the statistical ingredients developed by  \cite{cra-sabeti} in two directions. Most importantly, we will examine the performance of their Bayesian cubic spline estimator within an additive model framework. Secondly, we investigate the performance of the {\it cross validated marginal likelihood (CVML)} criterion that adapts  the seminal concept of cross-validation for marginal likelihood considered by \cite{geddy} to the conditional copula setting. 

In the next section we introduce the statistical model, describe  the computational algorithms needed for inference and the calculation of the CVML criterion.  Simulations and a real data analysis are discussed in Section 3. The paper closes with a discussion of future research directions.

\section{The Model}

In a regression setting we consider  the continuous bivariate outcome $Y_1,Y_2$ along with covariate $X \in \RR^p$. Marginally, each response $Y_i$, $i=1,2$  is modelled using a normal regression model. For a sample of size $n$, $\{(Y_{1j},Y_{2j},X_j): \; 1\le j \le n\}$,  where $X_j=(X_{j1},\ldots,X_{jp})^T$, we assume marginally
\beq
Y_{ij}\sim N(X_j^T\beta_i, \sigma_i^2), \; \forall 1\le i \le 2, \; 1\le j \le n,
\label{marg}
\eeq
and
joint density 
\begin{eqnarray*}
f(Y_{1j},Y_{2j}|X_j) &=& \prod_{i=1}^2{1\over \sigma_{i}} \phi \left ( {Y_{ij}- X_j^T\beta_i \over \sigma_i} \right ) \times \\
&\times&  c^{(1,1)}\left  \{
\Phi \left ( {Y_{1j}- X_j^T\beta_1 \over \sigma_1} \right),
\Phi \left ( {Y_{2j}- X_j^T\beta_2 \over \sigma_2} \right) \bigg |\theta(X_j)\right  \}, \; \forall 1\le j \le n,
\end{eqnarray*}
where $c^{(a,b)}(u,v|\theta) ={\partial^{a+b}C(u,v|\theta)/\partial u^a \partial v^b}$, for  all $0\le a,b \le  1$.

An important part of the model is the specification of $\theta(X)$. Many copula families have their parameter  $\theta$ restricted to a subset of $\RR$. In such cases we transform the parameter $\theta$ via a user-specified link function $g$ that maps the support of the copula parameter onto the real line and then we set  $g(\theta)=\eta(X)$, where 
$\eta:\RR^p \rightarrow \RR$ is the unknown {\it calibration function} we want to estimate. It is known that there is a one-to-one correspondence between the copula parameter $\theta(X)$ and the conditional Kendall's tau $\tau(X)=4{\rm E}\{H(Y_{1},Y_{2}|X)|X\}-1$ where the  mean is taken with respect to the joint conditional density of $(Y_{1},Y_{2})$ given $X$. Therefore, one can parametrize the model on the $\tau$ or $\theta$ scale. In this paper the inference is performed directly on the copula parameter calibration function for computational convenience. However, when goodness-of-fit measures are reported across different copula families, it is recommended to use the $\tau$ scale which is parametrization invariant \citep[see also discussion in][]{acy}.

When $p>1$ we  adopt an additive model  \citep{tibs-hast} for $\eta(X)$
\beq
\label{eta_all}
\eta(X)= \alpha_0+\sum_{i=1}^p \eta_i(X_i),
\eeq
where $\alpha_0 \in \RR$ and each $\eta_i: \RR \rightarrow \RR$ is specified using the flexible cubic spline model suggested by \cite{smith-kohn} in which
\beq
\label{eta_one}
\eta_i(X_i) =  \sum_{j=1}^{ 3} {\alpha}_j^{(i)} X_i^j +  \sum_{k=1}^{ K^{(i)} }{\psi}_k^{(i)} {(X_i - {\gamma}_k^{(i)})^3}_{+} \eeq
and  $a_+=\max(0,a)$. It is well known that the performance of   spline-based estimators are influenced by the location of  the knots $\gamma_k^{(i)}$. In our model this choice is automatic and  data-driven.  

A general remark is that in our implementations we assume that the covariates are independent random variables. In order to test this assumption when applying the method to real data, we have used tests based on the empirical copula process \citep{ind_gen_rem,koja} and  correlation of distances \citep{szek}.

The priors assigned to the parameters involved in the marginal models are:
\begin{eqnarray*}
\beta_i &\sim& N(0,\sigma_i^2 \II_p), \; \forall i=1,2\\
\sigma_i^2 &\sim & IG(0.1,0.1),\; \forall i=1,2.\\
\end{eqnarray*}
For the  parameters involved in the cubic spline  we follow the prior specifications used by \cite{cra-sabeti}. For each covariate $X_i$, we select a fixed value for the maximum number of knots,   $K_{\max}^{(i)}$. In the absence of additional information regarding which covariates are more likely to induce changes in $\eta$, we use the same $K_{\max}$ value for each $i=1,2,\ldots,p$. The  range spanned by the observed values of   covariate $X_i$ is divided into  $K_{\max}^{(i)}$ intervals of equal length, $I_1^{(i)},\ldots, I_{K_{\max}^{(i)}}^{(i)}$, and we assume  that each interval $I_k^{(i)}$ contains at most one knot. 
In order to complete the model specification, we introduce additional parameters $\{\zeta_k^{(i)}: \; 1\le k \le K_{\max}^{(i)}\}$, where for all $k \in \{1,\ldots,K_{\max}^{(i)}\}$
\[{\zeta}_k^{(i)} =  \left\{ 
 \begin{array}{l l}
   1 & \quad  \mbox{ if there is a knot } {\gamma}_k^{(i)} \in  I_k^{(i)},  \\
   0 & \quad  \mbox{ otherwise.}\\
  \end{array} \right.
 \]
 The model (\ref{eta_one}) becomes then
 \beq
\label{eta_two}
\eta_i(X_i)=\alpha_0+\sum_{j=1}^{ 3} {\alpha}_j^{(i)} X_i^j +  \sum_{k=1}^{ K_{\max}^{(i)}} \zeta_k^{(i)}{\psi}_k^{(i)} {(X_i - {\gamma}_k^{(i)})^3}_{+}
\eeq
and one can see that the number of non-zero terms in the sum depends on the values of $\zeta_1^{(i)},\ldots,\zeta^{(i)}_{K_{\max}^{(i)}}$.
For each $\eta_{i}$ we  construct a hierarchical prior for $\{\zeta_{1}^{{(i)}},\ldots, \zeta_{K_{\max}^{(i)}}\}$.  Specifically, if we let $|\zeta^{(i)}| = \sum_{k=1}^{K_{\max}^{(i)}} {\zeta}_k^{(i)}$ be the number of knots that are used in the  model for $\eta_i$ then 
 \beq
 p(|\zeta^{(i)}| \; | \; \lambda^{(i)}) \propto \frac{{\lambda^{(i)}}^{|\zeta^{(i)}|}}{|\zeta^{(i)}|!} \textbf{1}_{\{|\zeta^{(i)}| \leq K_{\max}^{(i)}\}},
 \label{zetas}
 \eeq
  i.e., $|\zeta^{(i)}|$ follows the right truncated Poisson distribution with parameter $\lambda^{(i)}$, and maximum value $K_{\max}^{(i)}$.
In addition, 
\begin{eqnarray*}
&&p(\zeta^{(i)} \; \big{|} \; |\zeta^{(i)}|) ={K_{\max}^{(i)} \choose |\zeta^{(i)}|}^{{-1}} ,   \\
&&p(\zeta^{(i)}|\lambda^{(i)}) = p(\zeta^{(i)} \; \big{|}\;  {|} \zeta^{(i)}|) p(|\zeta^{(i)}| \big{|}\lambda^{(i)}). 
\end{eqnarray*}
The form of $p(\zeta^{(i)} \; | \; |\zeta^{(i)}|) $  implies that, given a number of knots for the model, all configurations of intervals containing a knot are equally likely. The priors for all the parameters involved in the spline model for $\eta_i$ are chosen regardless of the type of outcome as
\begin{eqnarray}
&&\lambda^{(i)} \sim \mbox{Bin}(K_{\max}^{(i)}, p = 0.5),\nonumber  \\
&& \alpha_0 \sim \mathcal{N}(0, 10),\nonumber  \\ 
&&\alpha_j^{(i)} \sim \mathcal{N}(0, 10), \;  \forall 1\le j \le 3\nonumber\\
&&\psi_k^{(i)} \sim \mathcal{N}(0, 10), \;  \forall 1 \le k \le K_{\max}^{(i)}\nonumber \\
&&\gamma_{k}^{(i)} \sim \mbox{Unif}[I_{k}^{(i)}], \;  \forall 1\le k \le  K_{\max}^{(i)}.
\label{spriors}
\end{eqnarray}

\begin{figure}
\centering
\includegraphics[width=3.3in, height=2.7in, angle=270]{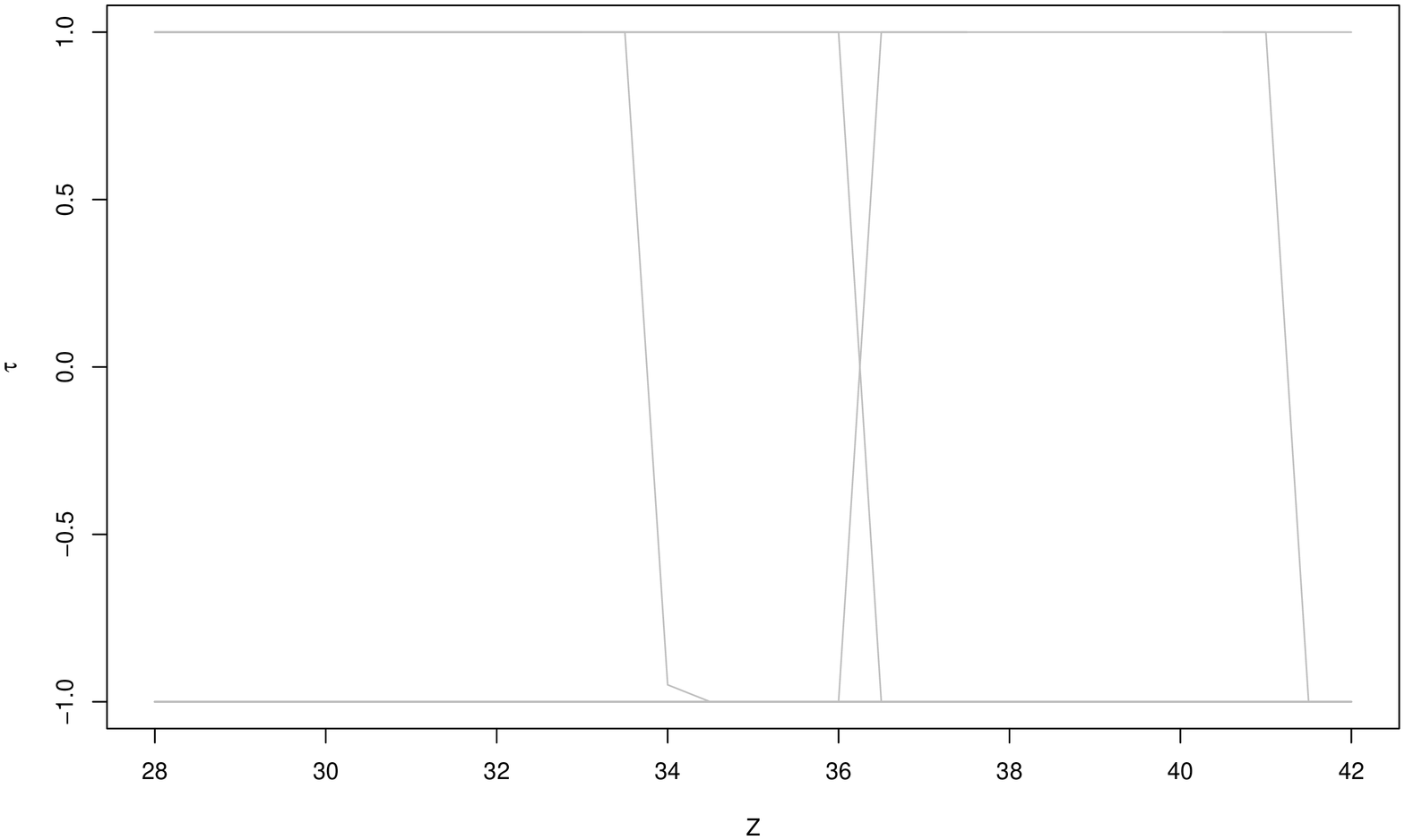}
\includegraphics[width=3.3in, height=2.7in, angle=270]{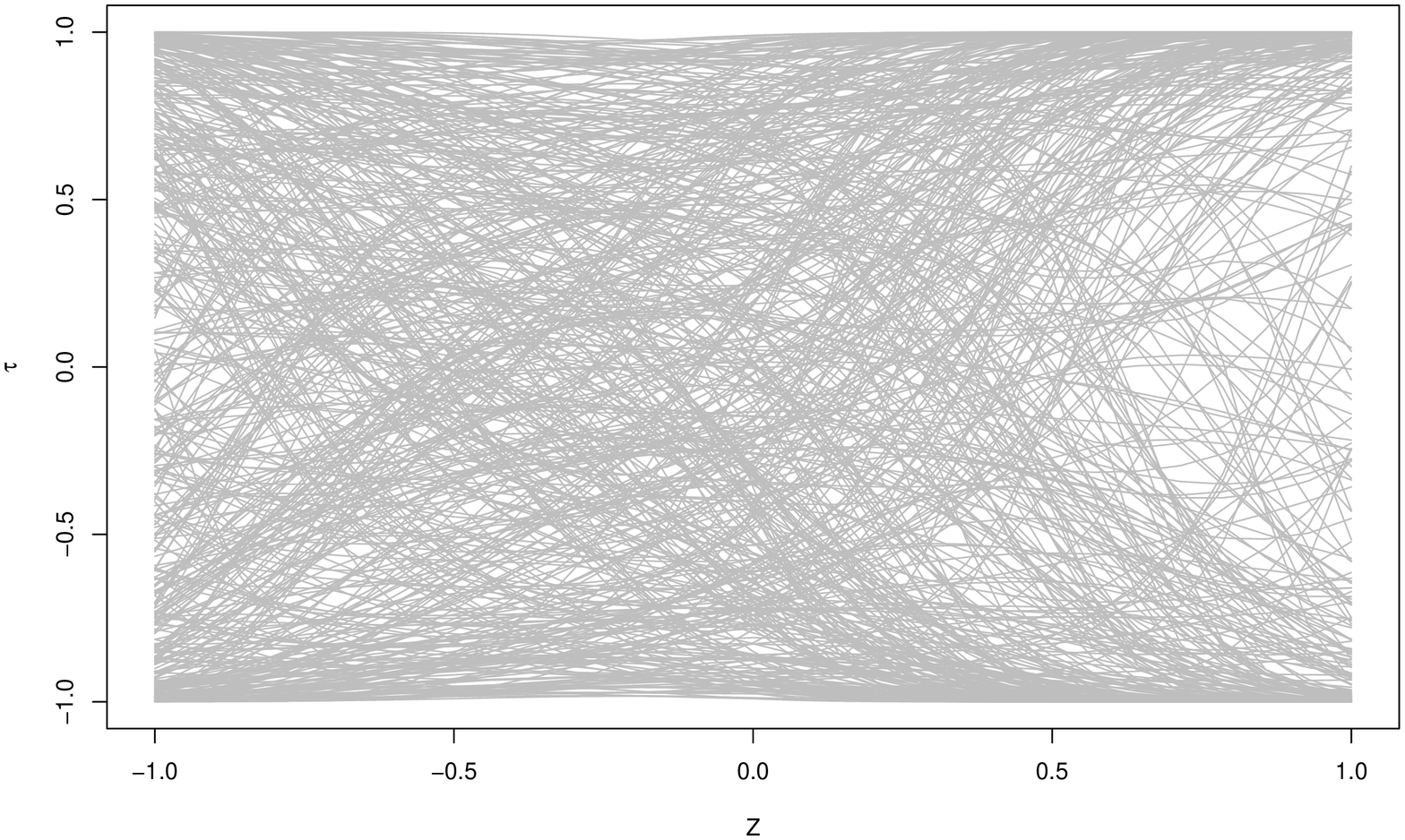}
\caption{500 realizations of $\eta_i(z)$  drawn from the prior distribution. The simulation setup is inspired by the real data example in Section 5: in the left panel $Z$ is uniform on (28,42) and  in the right panel $Z$ has been standardized using the transformation $h(Z)=(Z-35)/7$.}
\label{fig:sprior}
\end{figure}

Without additional information on the shape of $\eta_i$ we would like to be as vague as possible a priori. Note that the prior distributions given in equations (\ref{zetas}) and (\ref{spriors}) induce a prior distribution on the set of all possible maps $\eta_i : \RR \rightarrow \RR$. This prior is  too complex to characterize analytically, but easy to sample from.  Specifically, given a response index $i$, each sample of spline parameters $\{\zeta_k^{(i)}, \gamma_k^{(i)}, 
\psi_k^{(i)} : \; 1\le k\le K_{\max}^{(i)}\}$, $\{\alpha_j^{(i)} :  \; 1\le j \le 3\}$ and $\alpha_0$ from (\ref{zetas}) and (\ref{spriors}) will produce, when plugged into equation (\ref{eta_two}), a curve $\eta_i$.  If the priors used  are indeed not too informative about the shape of $\eta_i$ then we do not expect to see emerging any particular patterns. Our numerical experiments show that the prior is not too sensitive to changes in the values used  in  (\ref{zetas}) and (\ref{spriors}), but is sensitive to the covariate's range. In Figure \ref{fig:sprior} we show 500 maps $\eta_i(z)$ on the Kendall's tau scale where it has bounded range $[-1,1]$. The left panel illustrates the case where the covariate is uniform on the interval $(28,42)$ (the range was chosen to match the data example in Section 3.5) and the curves in the right panel are obtained after standardizing the covariate so that the new range is $[-1,1]$. When the   range for the covariate  is large the prior weight is assigned mostly to extreme dependence patterns where Kendall's tau is close to 1 or -1 for almost all values of $X$. Such priors are undesirable as they have the potential of biasing the inference. However, after standardizing the covariate, the prior bias seems to vanish. For this reason we recommend standardizing all covariates used in the conditional copula model.

\subsection{The Computational Algorithm}

If  $\omega$ is the vector of all the parameters involved in the model and $\dd$  are all the observed data, the posterior distribution $\pi(\omega|\dd)$ cannot be studied analytically due to its complicated form. Instead, we construct an Markov chain Monte Carlo (MCMC) algorithm to sample from $\pi(\omega|\dd)$. The form of the sampling algorithm follows the generic design of the Gibbs sampler \citep{gibbs-gelfand} in which every component  $\omega_{j}$ is updated by sampling from its conditional distribution  $\pi(\omega_{j}|\omega \backslash \omega_{j},\dd)$. Some of the components of the chain cannot be sampled directly from the conditional distribution, so a Metropolis-Hastings update is needed \citep[for details on using Metropolis-Hasting updates within the Gibbs sampler see, for instance,][]{cra-ros}.  The strategies used to update each parameter at step $t+1$ are described below. The super index $^{(t)}$ indicates the iteration step.

\begin{description}

\item[$\beta$'s:] Let $\BX \in \RR^{n\times p}$ be the matrix whose rows are $X_j^T$, $1\le j\le n$ and $\BY_1,\BY_2$ the response vectors, i.e. $\BY_i=\{Y_{ij}: \; 1\le j \le n\}$. If we had not considered the copula factor to account for the dependence between the outcomes,  the posterior conditional distribution of $\beta_1,\beta_2$ would have been available in closed form  
\beq
\tilde \pi_i(\beta_i|\dd,\sigma_i^{(t)})=\tilde \pi(\beta_i|\BX,\BY_i,\sigma_i^{(t)}) = n(\beta_i; \mu_i,\Sigma_i), \; i=1,2
\label{post_beta}
\eeq
where $n(x;a,b)$ is the density of a normal with mean vector $a$ and variance matrix $b$, and
\beqn
\mu_i&=& (\II+\BX^T \BX)^{-1} \BX^T \BY_i  \\
\Sigma_i&=& (\sigma_i^{(t)})^2 (\II + \BX^T \BX_i)^{-1} \nonumber , \;\; i=1,2.
\label{mu-sigma}
\eeqn

The update of each $\beta_i$ involves a mixture transition kernels. With probability $\lambda=0.8$ we update using  an Independent Metropolis (IM) transition kernel in which the proposal  distribution is $\tilde\pi_i(\beta_i)$ and with probability $1-\lambda=0.2$ we update using a Random Walk Metropolis (RWM) with a Gaussian proposal with mean at the current value of $\beta_i$ and variance chosen so that the acceptance rate is between 20-30\%.  

\item[$\sigma$'s:] Once again, without the copula component of the likelihood, the posterior conditional distribution of $\sigma_i$, given the data and $\beta_1,\beta_2$, is available in closed form
 
\begin{eqnarray}
&&\tilde \pi (\sigma_i|\dd,\beta_i^{(t+1)})= \tilde \pi (\sigma_i|\BX,\BY_i,\beta_i^{(t+1)})= \nonumber\\
&=&IG\left(0.1+{p +n \over 2},0.1+{(\beta_{i}^{(t+1)})^{T}\beta_{i}^{(t+1)} +(\BY_{i}-\BX\beta_{i}^{(t+1)})^{T}(\BY_{i}-\BX\beta_{i}^{(t+1)}) \over 2} \right), \; i=1,2.\nonumber
\label{post_sigma}
\end{eqnarray}

The updates are made according to an IM kernel in which the proposal density is $\tilde \pi (\sigma_i|\BX,\BY_i,\beta_i^{(t+1)})$ for each $i=1,2$. 
The updating steps for $\beta$ and $\sigma$ lead to faster mixing compared to those defined
in \cite{cra-sabeti} where only RWM updates were used, because the IM transition kernel allows the chain to jump around the target space and reduces autocorrelation.

\item[$\alpha$'s:] Because there is no range restriction for each $\alpha_k^{(i)}$  and no direct sampling strategy is possible,   we use the RWM-within-Gibbs with  proposal variance tuned so that the acceptance rates are between 20-40\%. 

 \item[$\zeta$'s:] The updates are performed using the Metropolis-within-Gibbs strategy for the {\it entire} latent variable vector $\vec\zeta^{(i)}=(\zeta_1^{(i)},\ldots,\zeta_{K_{\max}}^{(i)})$. 
 For updating $\vec{\zeta}^{(i)}$ we use two type of moves: we either add/delete a component (i.e. transforming a zero component into a one or vice-versa) or swap two components.  We choose with probability half to either add/delete a component chosen or to permute two components of $\vec\zeta$ that are selected at random. Each proposed move is accepted or rejected based on a Metropolis-Hastings rule.
 
\item[$\psi$'s:] If $\zeta_{k}^{(i)}=1$ we use the RWM-within-Gibbs strategy to update  $\psi_k^{(i)}$ using proposals tuned so that the acceptance rates are between 20-50\%. If $\zeta_{k}^{(i)}=0$, $\psi_k^{(i)}$ is updated using a random draw from its prior distribution that is automatically accepted.

\item[$\gamma$'s:]  If $\zeta_{k}^{(i)}=1$  we  use an IM update for $\gamma_k^{(i)}$ using as proposal the prior distribution of $\gamma_k^{(i)}$.  If $\zeta_{k}^{(i)}=0$ then the next state  $\gamma_{k}^{(i)}$ is sampled from its prior and automatically accepted. 

\item[$\lambda$:]  For $\lambda$ we  use an IM update with proposal distribution equal to the prior, i.e. Bin$(0.5,K_{\max})$.

\end{description}

\subsection{Cross Validated  Marginal Likelihood Model Selection}

The cross-validated, pseudo marginal likelihood (CVML) criterion of 
\cite{geddy} is used to compare the predictive power of various models considered. Denote $\M$ such a generic model, characterized by regression parameters $\{\beta_i,\sigma_i:\; i=1,2\}$ corresponding a  subset of covariates, $\BX$, and all the spline parameters involved in modelling the calibration function $\eta(\BX)$. Denote the parameters in the model  $\omega$,  the data is $\dd$ and for each $1\le j \le n$, $\dd_{-j}$ denotes the remaining data after we have removed the covariates  and responses pertaining to the $j$th item, $(Y_{1j},Y_{2j},X_j)$. 
A selection criterion based on the  CVML will choose the model $\M$ that maximizes the sum
\beq
CVML(\M)=\sum\limits_{j=1}^{n}\log p(Y_{1j},Y_{2j}|\dd_{-j},\M).
\label{eq:cvml}
\eeq 
One can see from \eqref{eq:cvml} that the CVML criterion favours models that exhibit  good average  predictive power.  The average is taken with respect to the parameters in the model so (\ref{eq:cvml}) is a function of the observed data only. From a Bayesian standpoint the computation of the criterion would be impractical  if we were to proceed by performing separately $n$ data  analyses, one for each sample of size $n-1$.  However, the following simple derivation can be used to compute $CVML(\M)$ from  a single Bayesian analysis of the {\it whole} data \citep[see also][]{hanson}. We have
\beqn
 E[p(Y_{1j},Y_{2j}|\omega)^{-1}]  &=&{1 \over p(\dd|\M)} \int {p(\dd| \omega,\M) p( \omega|\M) \over p(Y_{1j},Y_{2j}| \omega,\M)} d\omega={1 \over p(\dd|\M)} \int p(\dd_{{-j}}| \omega,\M) p(\omega|\M)d\omega = \nonumber \\
&=&{p(\dd_{-j}|\M)\over p(\dd|\M)}={1 \over p(Y_{1j},Y_{2j}|\dd_{-j},\M)},
\label{der-cvml}
\eeqn
where the first expectation is taken with respect to the posterior distribution of all parameters in the model, $\pi(\omega|\dd,\M)=p(\dd|\omega,\M)/p(\dd|\M)$.
Based on (\ref{der-cvml}) we deduce that a Monte Carlo estimator of (\ref{eq:cvml}) is 
\beq
\widehat{CVML}(\M)= \sum_{j=1}^n -\log \left [ {1\over M} \sum\limits_{m=1}^{M} p(Y_{1j},Y_{2j}|\omega^{(m)},\M)^{-1} \right ],
\label{est-cvml}
\eeq
where $\omega^{(1)},\omega^{(2)},\ldots,\omega^{(M)}$ are draws from the posterior distribution $\pi(\omega|\dd,\M)$ obtained via the MCMC algorithm described in the previous section.

\section{Simulations}

The simulation study provides information about the average errors incurred when implementing the proposed estimation approach   and  illustrates the  performance of the CVML criterion when it is used to 
select the copula family and the influential covariates in model (\ref{eta_all}).

\subsection{Simulation Details} 

We have generated  data using the Clayton copula using either a univariate or a bivariate calibration function.  Marginally, the outcomes follow the distributions defined by the linear models specified in (\ref{marg}). All covariate values are independently sampled from the  $\ru[0,1]$ distribution. For the dependence structure we have considered two nonlinear calibration functions $\eta_{S1},\eta_{S2}$  defined as
$$\eta_{S1}(x)=\log[4.5 - 1.5\sin (\pi x)],$$
and
$$\eta_{S2}(x_1,x_2)=  \log[4.5-\sin(x_1) - \sin(x_2)].$$

Under scenario {\bf S1} we simulate data using only one covariate so the true calibration function is  $\eta_{S1}$ and under  scenario {\bf S2} we generate data using the calibration
$\eta_{S2}$.
Marginally, under  {\bf S1} and {\bf S2}, each response variable is linked to, respectively, one or two covariates via a linear model with Gaussian errors, as specified in (\ref{marg}).  

Each  analysis has been independently replicated 50 times for samples of size $n=450$. We kept $K_{max}=4$ fixed throughout the simulation study.   The MCMC sampler was run for 10,000 iterations and the first  3000 samples were discarded as burn-in. The simulation parameters used in the MCMC samplers were selected so that the acceptance probabilities are between 20-40\%. The copula model data was generated using the {\tt{copula}} library within R. The main steps of the MCMC sampler were implemented in C++ with the results processed in R.

\subsection{Estimation of the Calibration Function}

In this section we present plots and measures  of the goodness-of-fit for  the estimating procedure proposed in this. We focus on scenario {\bf S2} which is more challenging to fit.

To provide a graphical illustration of the fit, in Table \ref{fig:thetaslice} we show one-dimensional slices in the true surface (black line), the estimated surface (red line) and the two surfaces delimitating the pointwise 95\% credible region (green lines). The slices are obtained when one of the two covariates 
is fixed at values in the set $\{-0.75,-0.25,0.25,0.75\}$.  We observe that the credible  bands grow wider near the boundaries of the covariate range and the bias gets also bigger when one of the covariate is closer to 1 or -1.

Table \ref{fig:theta} contains the trace plots, the autocorrelation plots (up to lag 200) and the histograms of the posterior sample realizations for $\theta(-0.25,0.75)$, $\theta(0.75,-0.25)$  and ${\theta}(0.75,0.75)$. 
In general, the ACF plots and the trace plots look similar. In the histograms, the red line shows the true value of the calibration function. We observe that the samples for $\theta(0.75,0.75)$ are further  from the true value when compared to the samples for  $\theta(0.75,-0.25)$. This is consistent  with our previous  observation concerning the fit when covariate values are close to the boundary.

We also look at the model estimates for the normal regression parameters. Table \ref{fig:beta} shows the trace plots, the autocorrelation plots and the histograms  obtained from posterior samples corresponding to the linear regression model for the first outcome, $\beta_{11}$ and  $\beta_{12}$, and the residual standard deviation $\sigma_1$.  The parameters used in the 
second response regression yield similar plots.

The red line in the histograms represents the true value of the parameters. 
Although the ACF seems to be high for these estimates, the histograms suggest that the samples provide  good estimates for  the marginal models parameters.

For a more global summary, we approximate numerically the integrated variance (IVAR), squared bias (IBias$^2$),  and mean squared error (IMSE) using a grid of  400 equidistant points in the covariate space. The values are reported in Table \ref{tb:imse}. When comparing these measures across the two simulation scenarios, we notice a significant increase in the bias when the number of covariates is increased. This is not surprising since the sample size is kept constant, but we fit a  significantly more complex  model  under scenario {\bf S2} than under {\bf S1}.

\subsection{Copula Selection}

We explore the performance of the CVML criterion for choosing the correct copula family. Specifically, we fit the generated data using Clayton, Frank and Gumbel copula families. In Table \ref{tb:cvmls} we report the percentage of correct decisions computed from 100 replicates. It can be noticed that there is a small decrease in 
accuracy for scenario {\bf S2} compared to {\bf S1} which is not 
surprising given that the former model is more complex than the latter.

\subsection{Variable Selection}

We have also examined the performance of CVML in selecting the covariates to be included in the model. We focused on data generated under scenario {\bf S2} and we fitted them using models with 1, 2, or 3 covariates. In all simulations results reported in this section we have used the correct Clayton copula to formulate the model.

If we denote $\M_i$ as the model with the first $i$ covariates included, then we see  from the box plots shown in Table \ref{cvml} that $CVML$ always selects $\M_2$ over $\M_1$ or $\M_3$. The difference in CVML values is larger between $\M_2$ and $\M_1$ than between $\M_2$ and $\M_3$, which is natural given the criterion's connection to the models  predictive power.

\subsection{Application to the Twin Birth Data}

The additive model approach is  applied to a subset of the Matched Multiple Birth Data Set. The data containing all twin births in the United States from 1995 to 2000 enable detailed investigation of twin gestations. We consider the twin live births in which both babies survived their first year of life with mothers of age between 18 and 40. Of interest is the dependence between the birth weights of twins (in grams), denoted by BW1 and BW2, respectively. We consider a random sample of 450 twin live births and investigate the effect of two covariates, gestational age (GA) and maternal age (MA), on the dependence between BW1 and BW2.

We compare the model $\M_{1}$ in which the GA is the only  covariate considered and  model  $\M_{2}$ in which  GA and MA are the included covariates. We also compare three analyses based on three parametric copula families: Clayton, Frank and Gumbel.  For each copula family we compute the CVML criterion for the models with both covariates (GA and MA) included. The results shown in the first row of Table \ref{twin-cvml} suggest that the Frank copula is more suitable for analyzing the data.  

Under the Frank copula, model $\M_{1}$ is preferred with a CVML  value of -5569.4 compared to -7683.7 obtained for $\M_{2}$. After deciding that $\M_{1}$ is preferred, we compare again the fit for $\M_{1}$ under each of the three copulas, and the results are shown on the second row of Table \ref{twin-cvml}.
This finding is concordant with the single covariate analysis of \cite{acy}.

\section{Conclusions and Future Work}

We propose Bayesian inference for the conditional copula model in a regression context with multiple  covariates. We implement  spline approximation within the additive model framework and propose a  model selection criterion which selects the model with the best predictive power. 

The simulations show that the efficiency of the method decreases as the dimension of the covariate vector increases and we would like to explore theoretically the rate of the  decay.  It is conceivable that when the number of covariates grows large, the approach proposed here may become too computationally expensive and simpler formulations of the calibration function and improvements of the MCMC algorithm needed to sample the posterior distribution are worth investigating.

\section*{Acknowledgment}

This work was supported by an individual NSERC of Canada research grant.

\bibliographystyle{ims}
\bibliography{superref}

\newpage

\begin{table}
\centering
\begin{tabular}{|c|c|c|}
\hline
Scenario $\backslash$ Copula&  Frank & Gumbel\\
\hline
{\bf S1} & 100  & 98\\
\hline
{\bf S2} & 96 &  94\\
\hline
\end{tabular}
\caption{PPerformance of CVML in selecting the correct Clayton family over Frank or Gumbel family under scenarios {\bf S1} and {\bf S2}. The numbers in the table represent the percentage of correct decisions.}
\label{tb:cvmls}
\end{table}

\begin{table}[ht]
	\centering
	\begin{tabular}{c}
		\includegraphics[height = 10cm,angle=270]{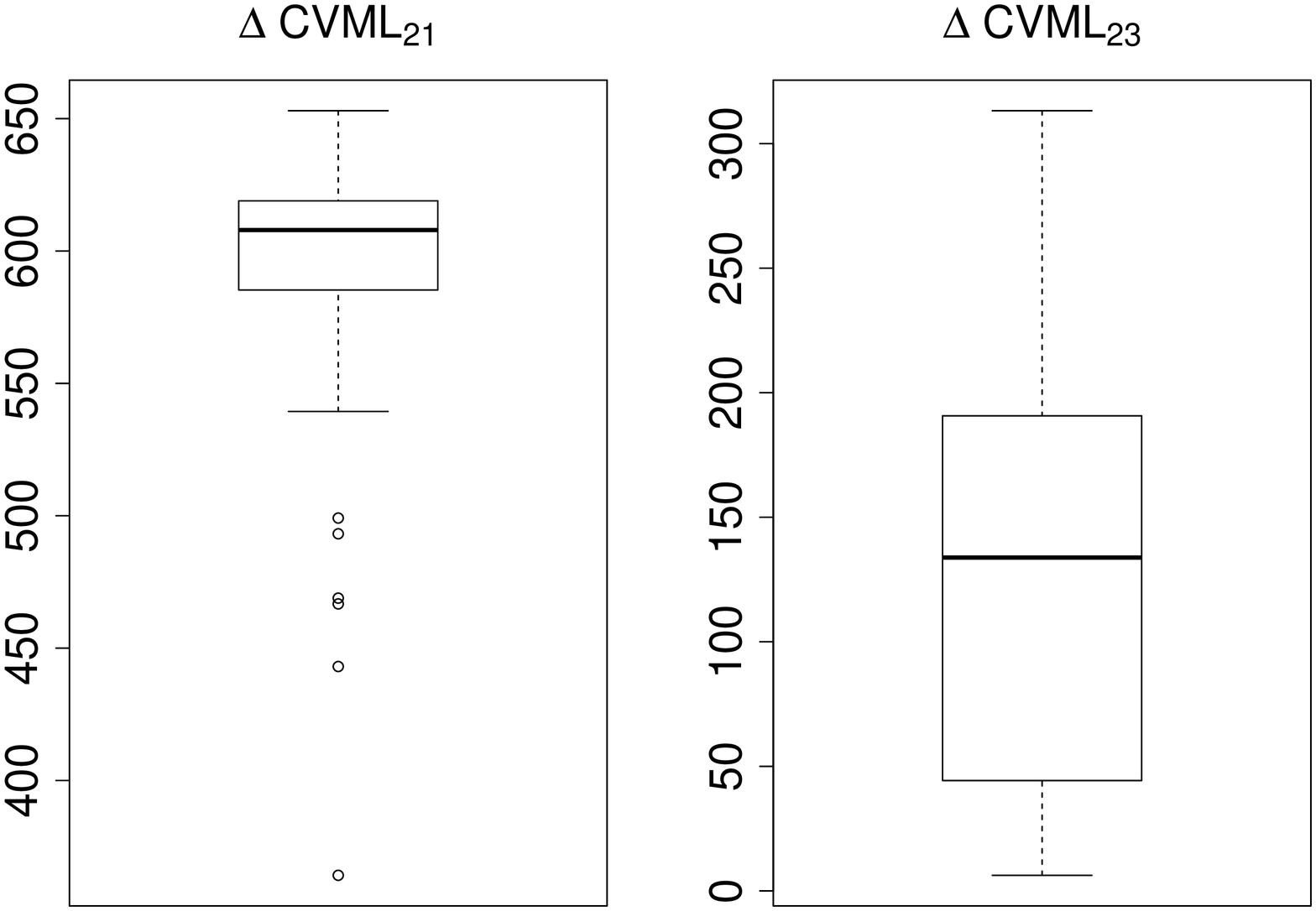}
	\end{tabular}
	\caption{Comparison of  the  CVML criterion values  for models with 1,2, or 3 covariates. Left panel: box plot of 100 independently replicated values of the difference $CVML(\M_2)-CVML(\M_1)$. Right panel: box plot of 100 independently replicated values of the difference $CVML(\M_2)-CVML(\M_3)$.  }
\label{cvml}
\end{table}

\begin{table}[ht]
	\centering
	\begin{tabular}{| c | c |}
		\hline
		Z$_1$ is fixed & Z$_2$ is fixed \\
		\hline 
		\includegraphics[height = 7cm,angle=270]{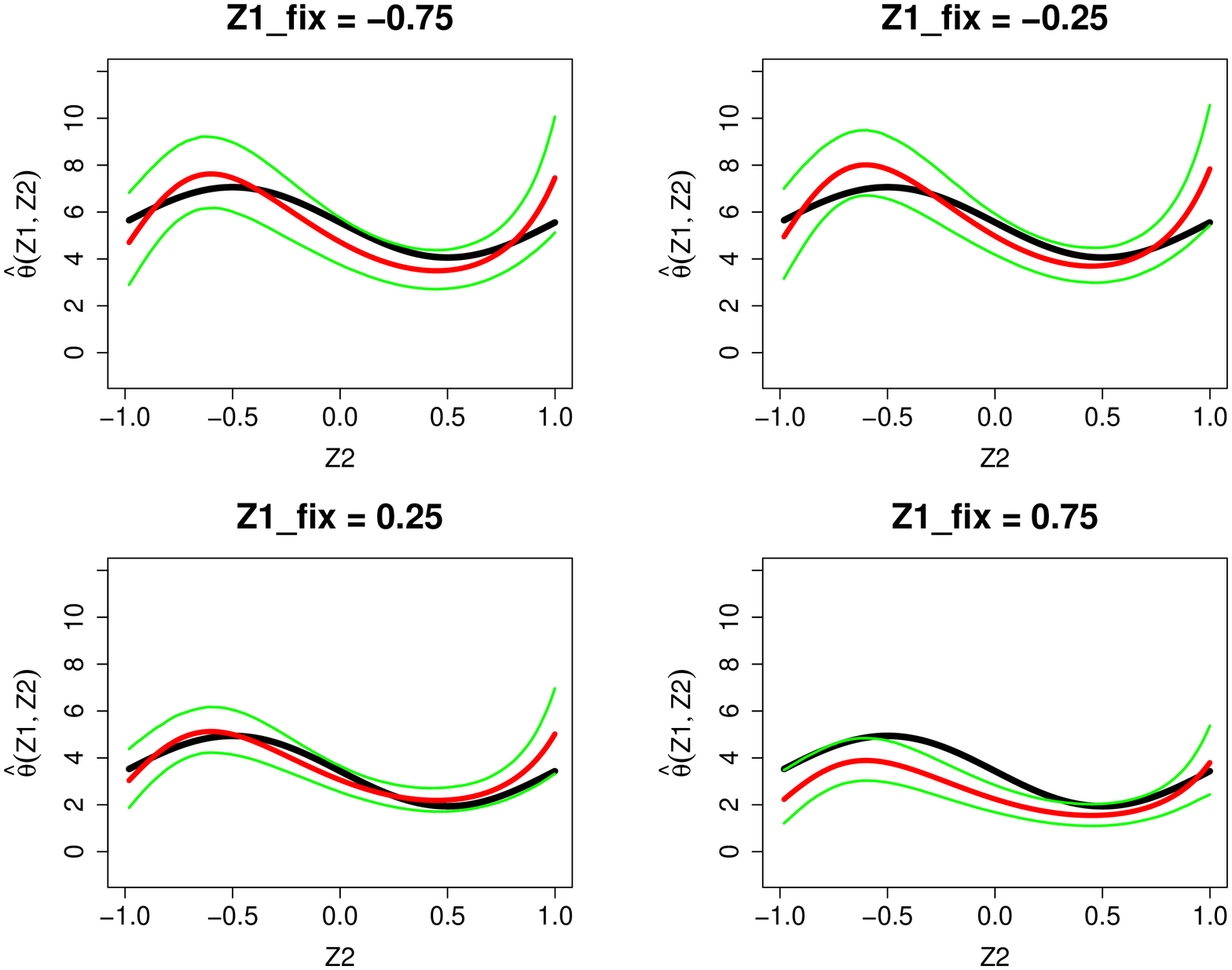} & \includegraphics[height = 7cm,angle=270]{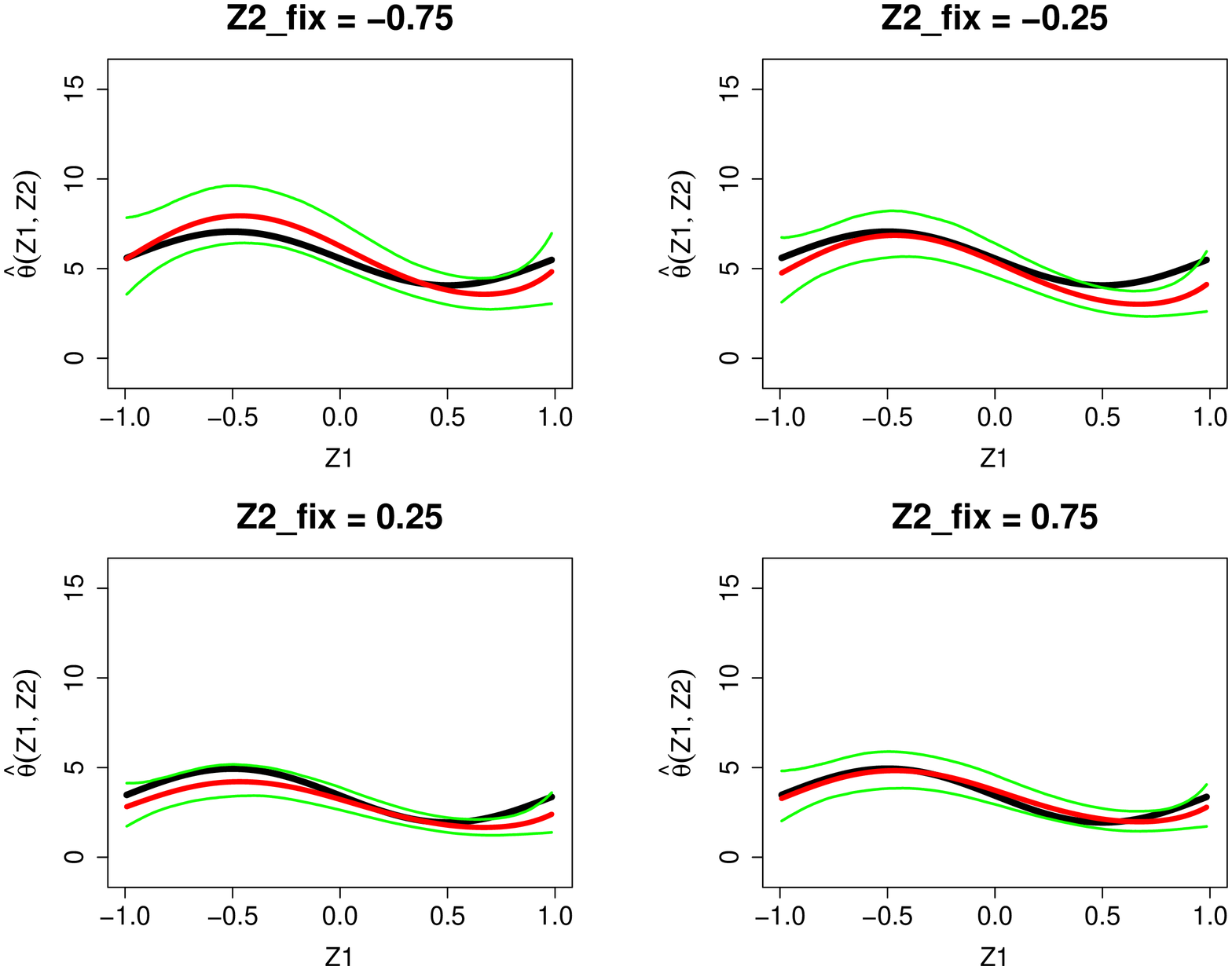}\\
		\hline

	\end{tabular}
	\caption{One-dimensional projections of the true calibration surface (black), the estimated surface (red) and confidence bands (green) produced under scenario {\bf S2}.}
	\label{fig:thetaslice}
\end{table}

\begin{table}[ht]
	\centering
	\begin{tabular}{| c | c |c|c|}
		\hline
		Scenario & 
		IBias$^2$ & IVAR & IMSE\\
		\hline
		{\bf S1} & 0.061 & 0.433 & 0.494\\
		{\bf S2} &0.132 & 0.515& 0.647 \\
		\hline
	\end{tabular}
	\caption{Performance of the estimation procedure under scenarios {\bf S1} and {\bf S2}.}
	\label{tb:imse}
\end{table}

\begin{table}[ht]
  \centering
  \begin{tabular}{|c| c |c | c |}
		\hline
		& $\hat{\theta}(-0.25,0.75)$ & $\hat{\theta}(0.75,-0.25)$  & $\hat{\theta}(0.75,0.75)$ \\
		\hline
    Trace Plots & \includegraphics[height = 4.5cm,angle=270]{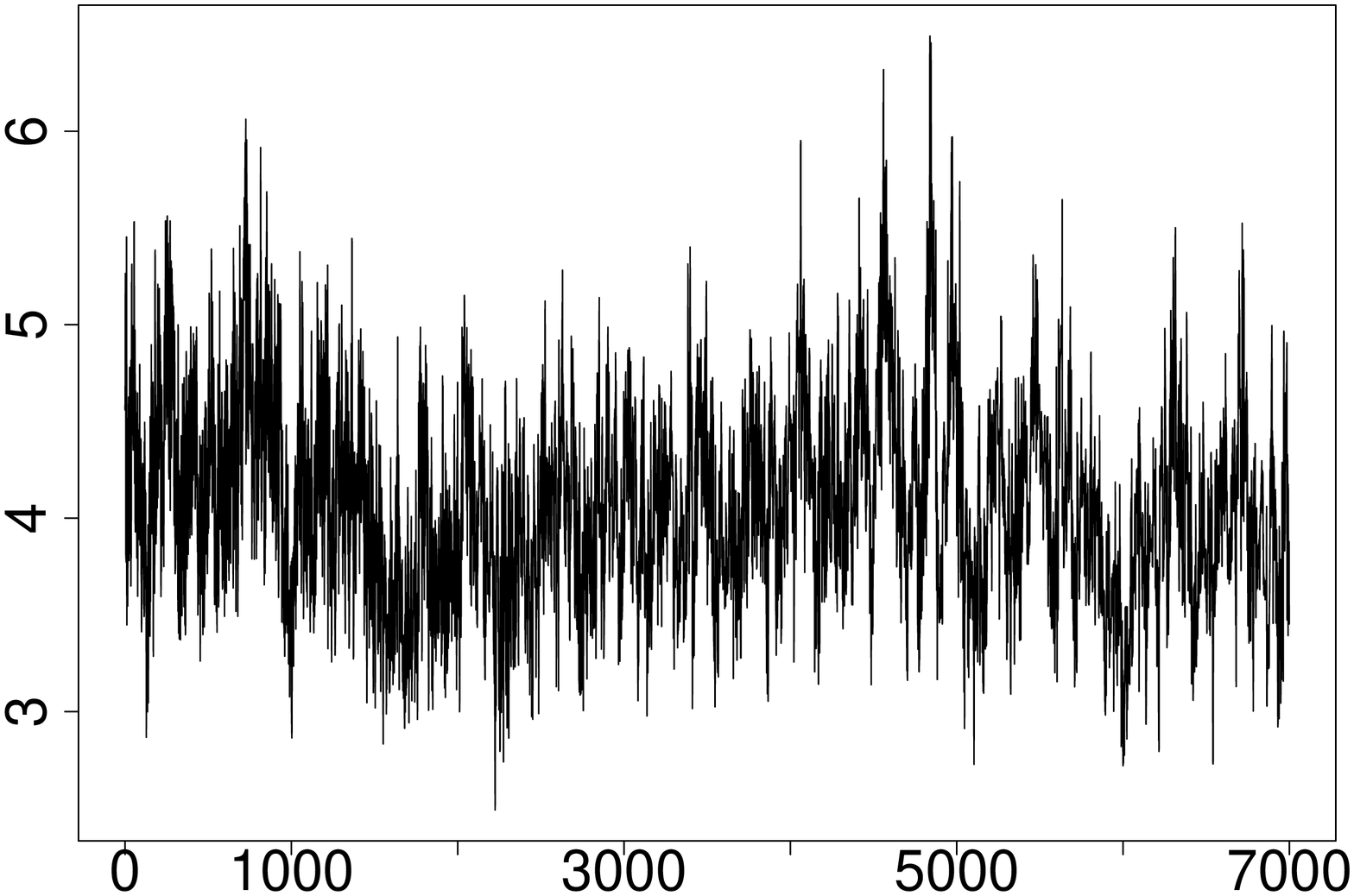}
      & \includegraphics[height = 4.5cm,angle=270]{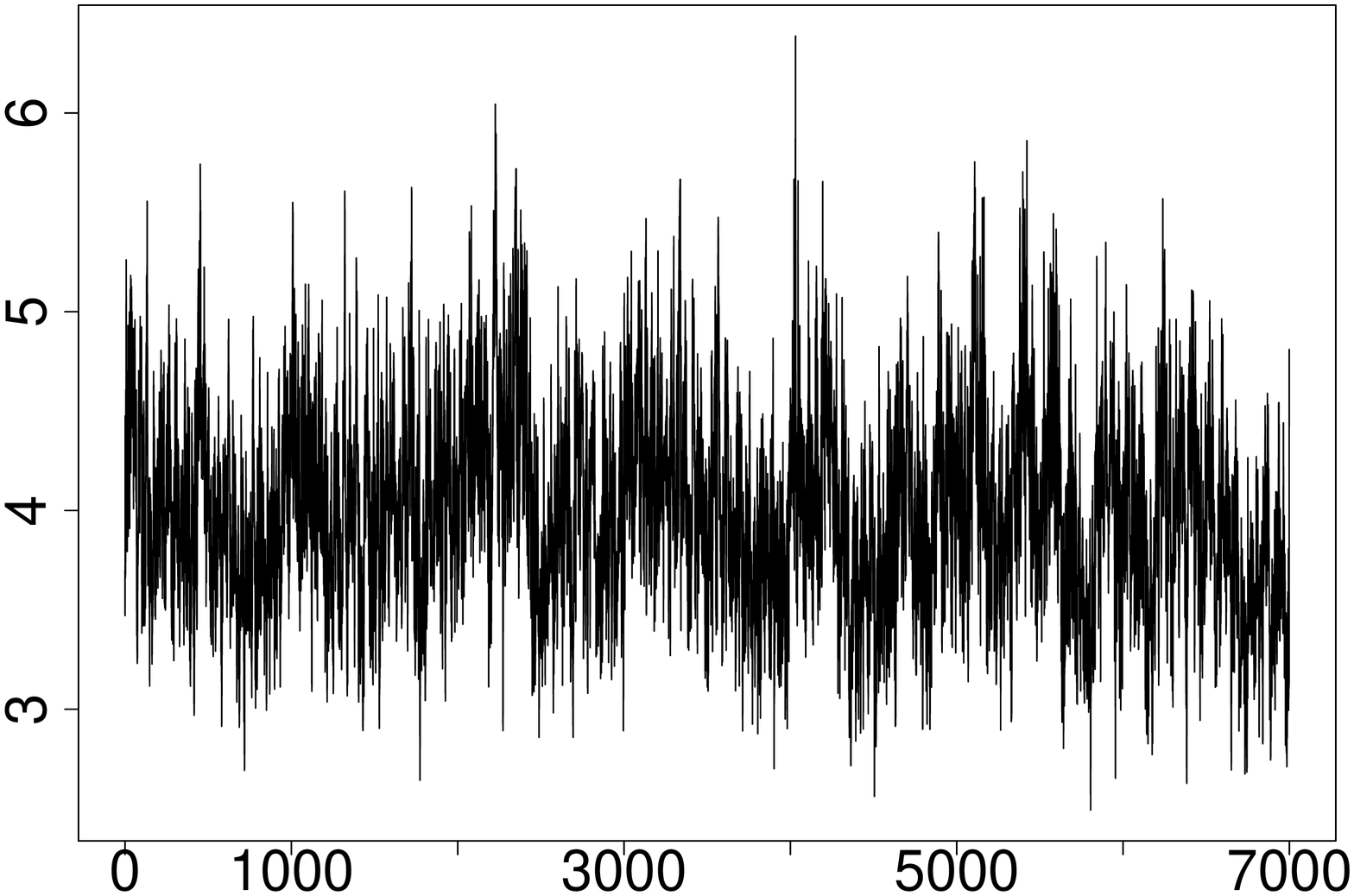} & \includegraphics[height = 4.5cm,angle=270]{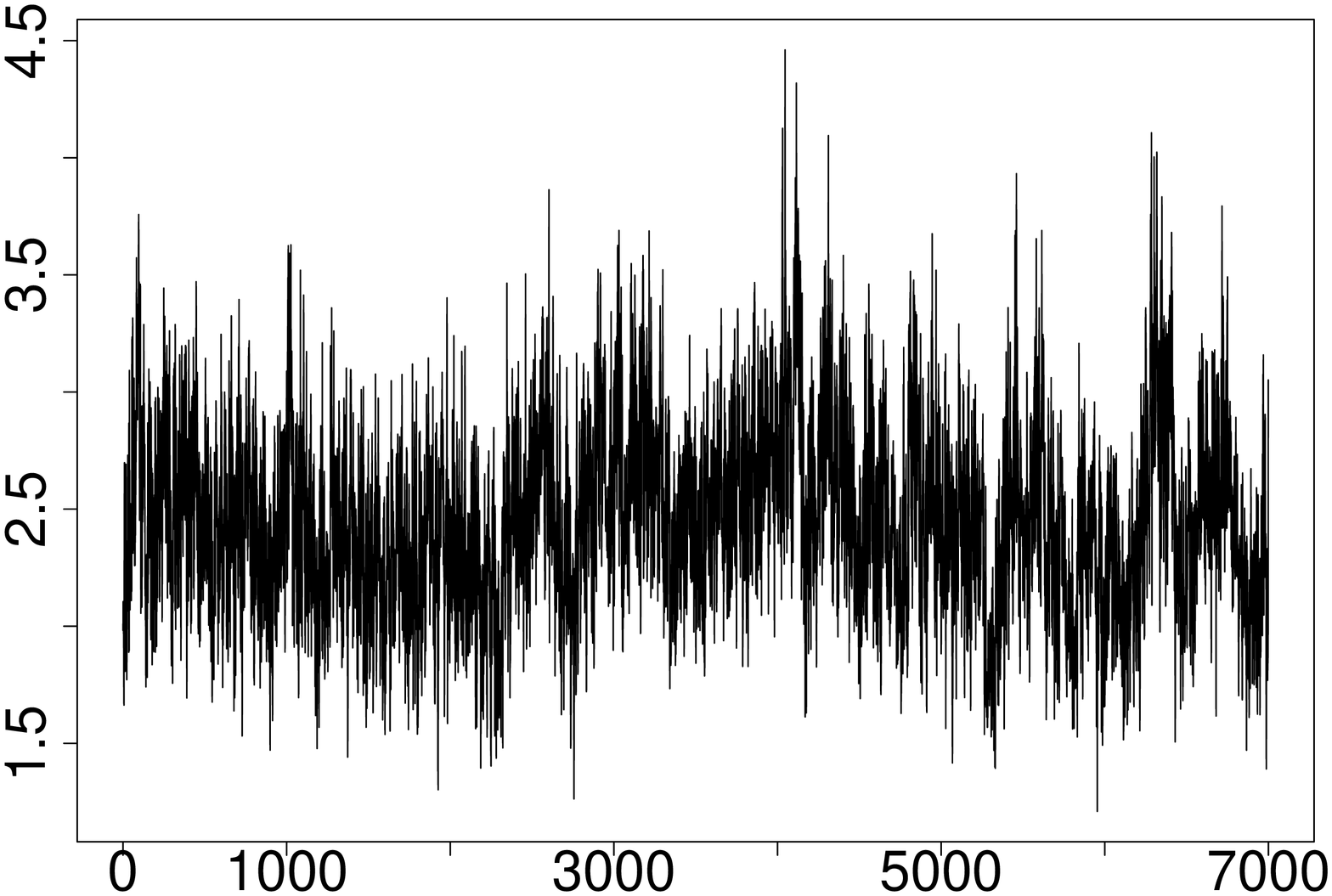} \\
		\hline
    Acf Plots & \includegraphics[height = 4.5cm,angle=270]{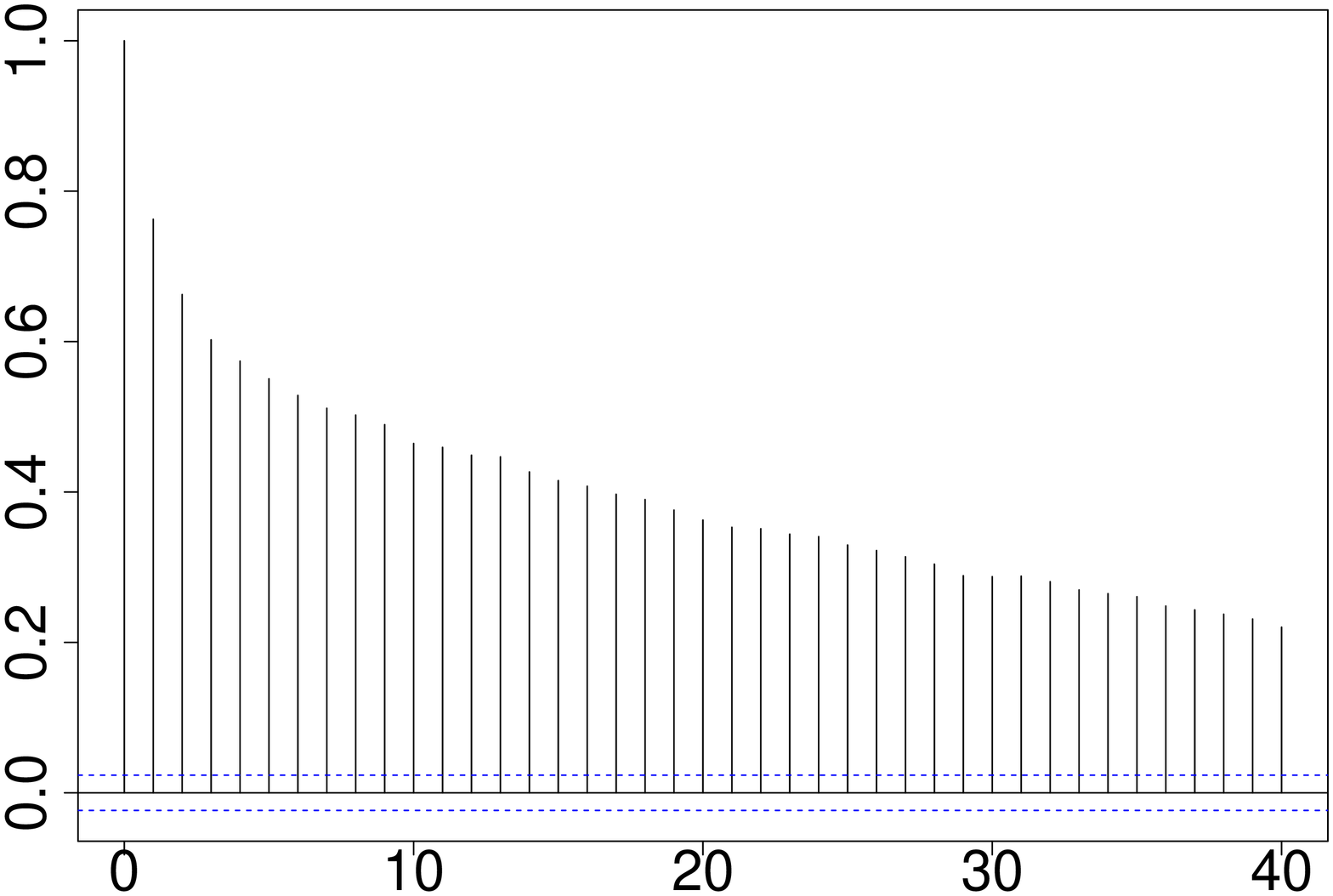}
      & \includegraphics[height = 4.5cm,angle=270]{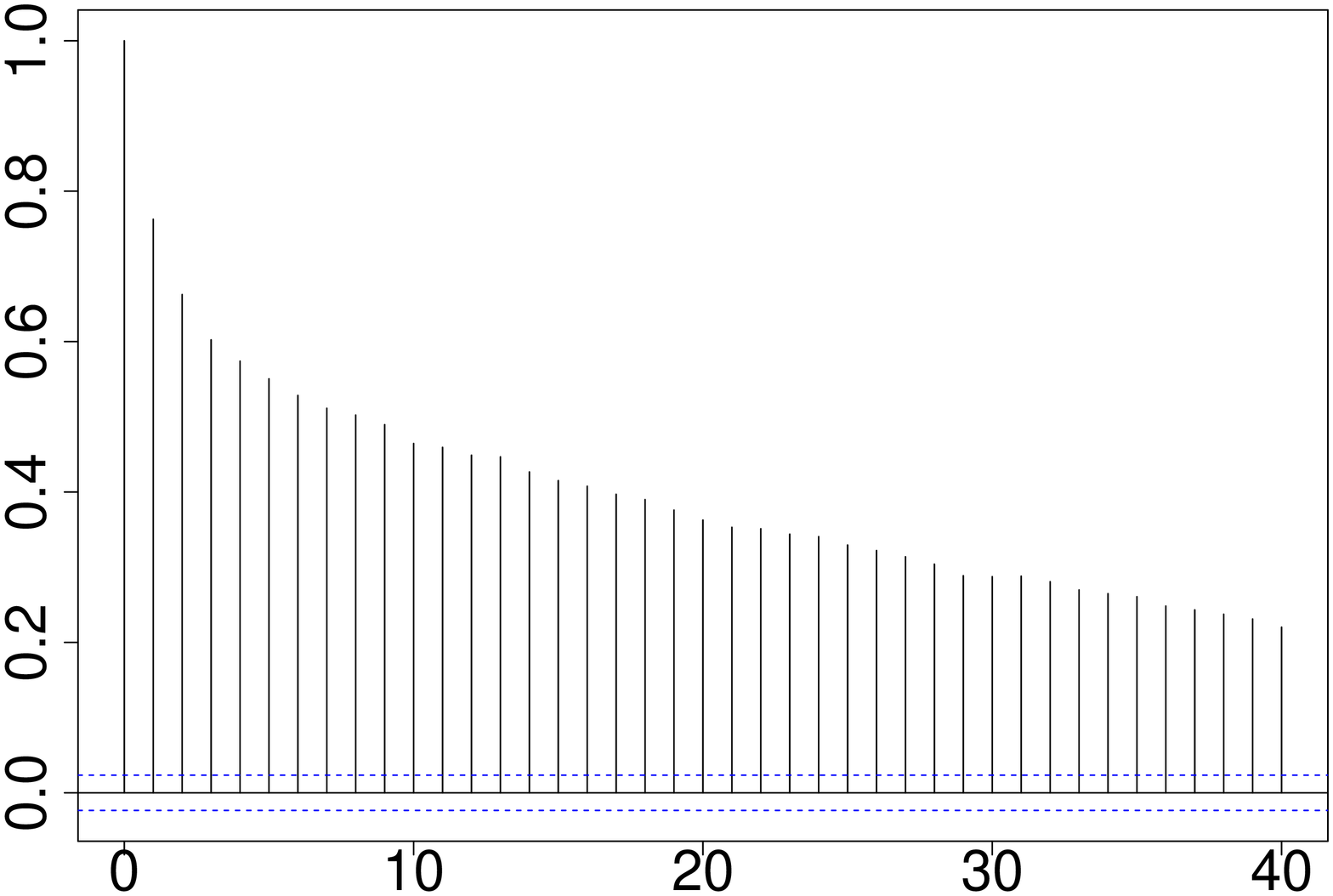} & \includegraphics[height = 4.5cm,angle=270]{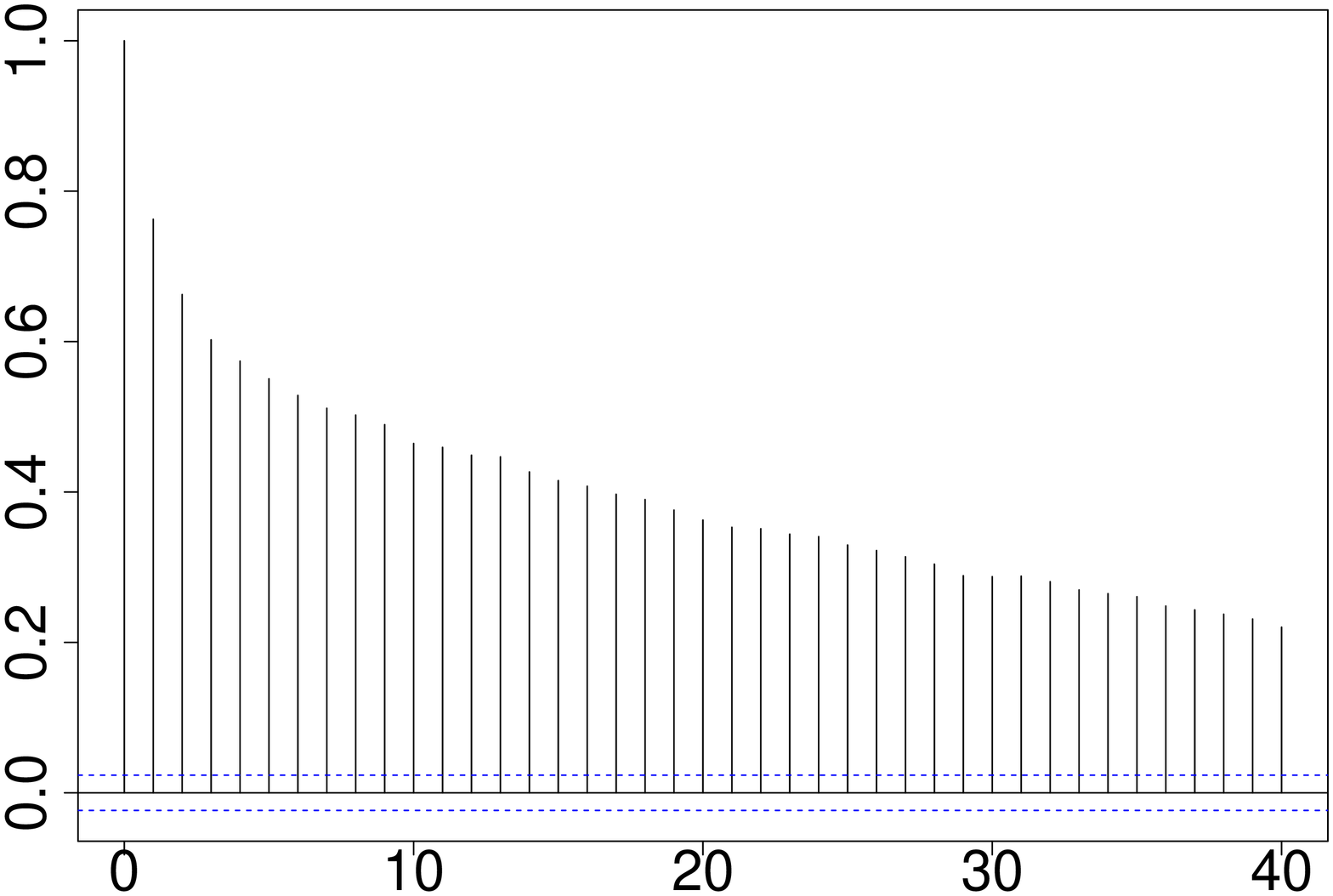} \\
		\hline
    Histogram & \includegraphics[height = 4.5cm,angle=270]{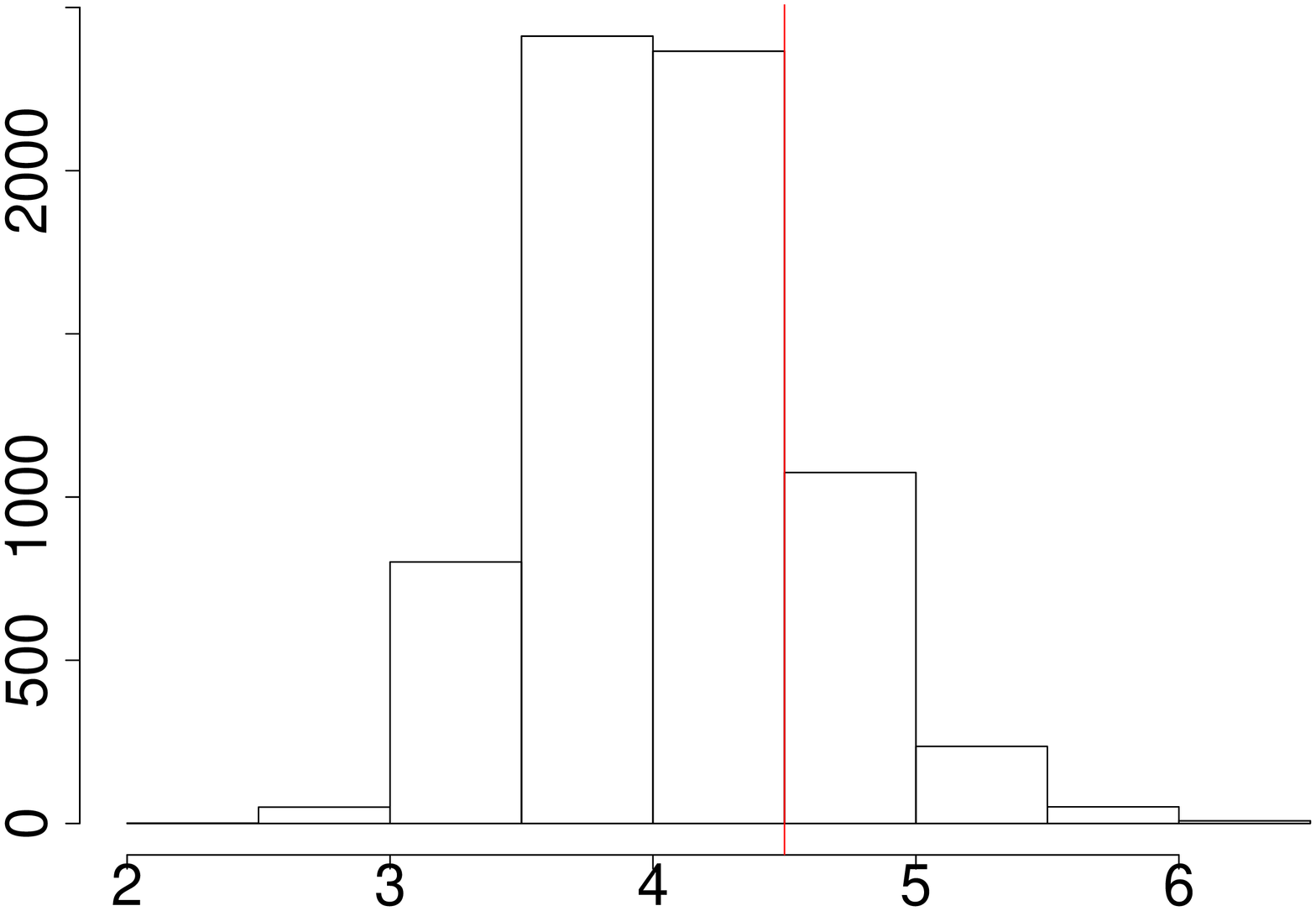}
      & \includegraphics[height = 4.5cm,angle=270]{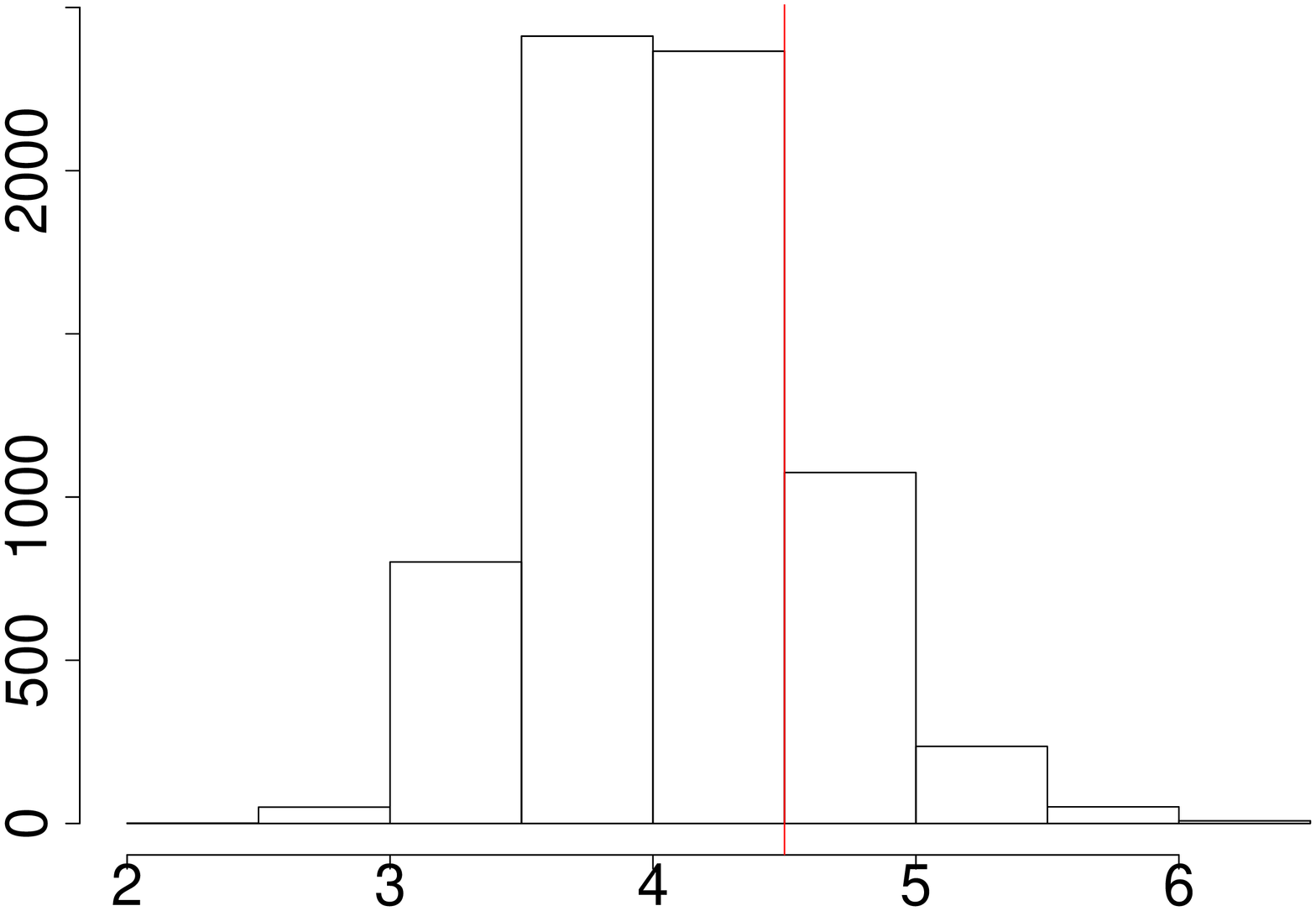} & \includegraphics[height = 4.5cm,angle=270]{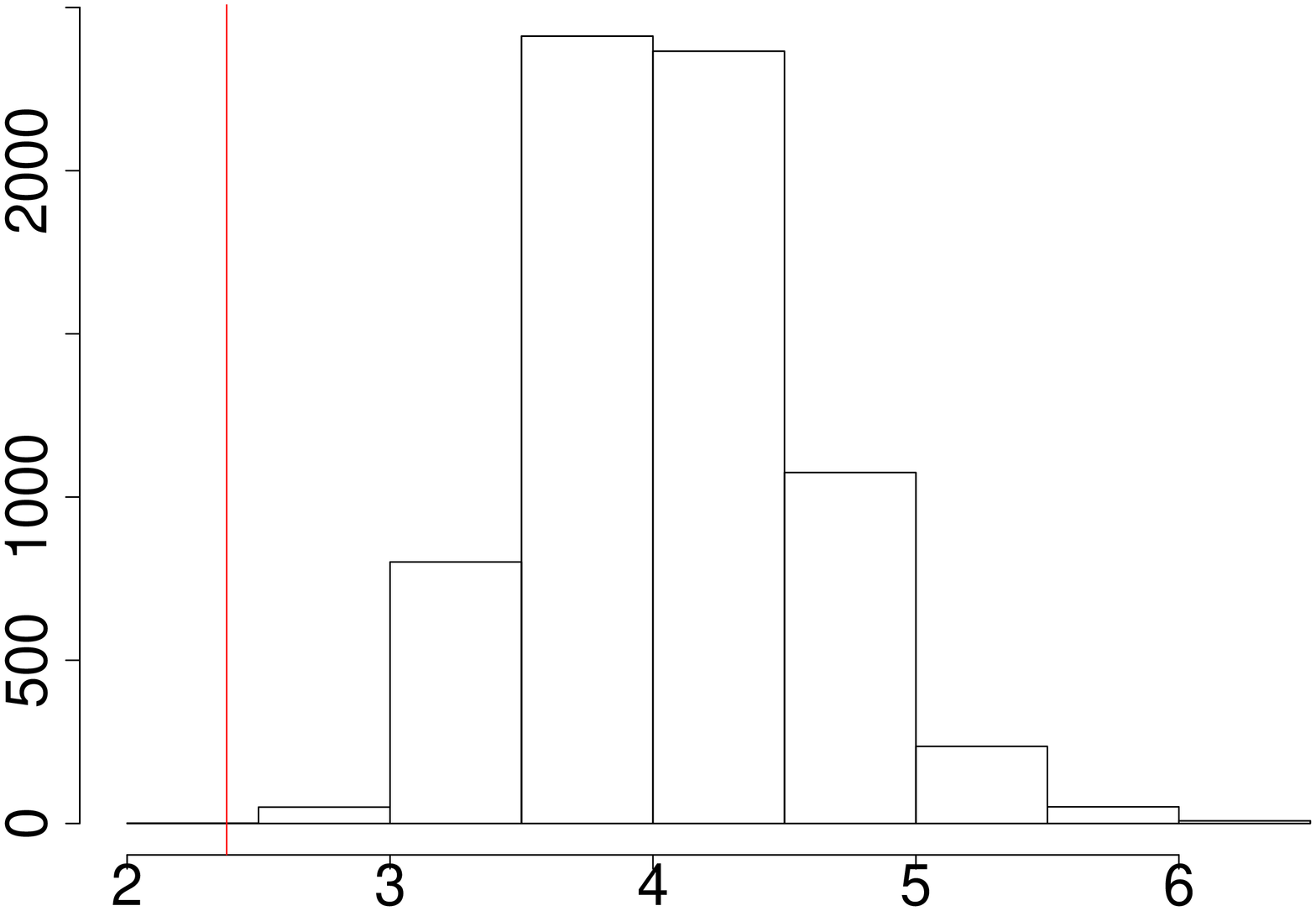} \\
		\hline
  \end{tabular}
  \caption{Plot details for $\theta$ in the 10th run.}\label{fig:theta}
\end{table}

\begin{table}[ht]
  \centering
  \begin{tabular}{|c| c |c | c |}
		\hline
		& $\beta_{11}$ & $\beta_{12}$ & $\sigma_1$ \\
		\hline
    Trace Plots & \includegraphics[height = 4.5cm,angle=270]{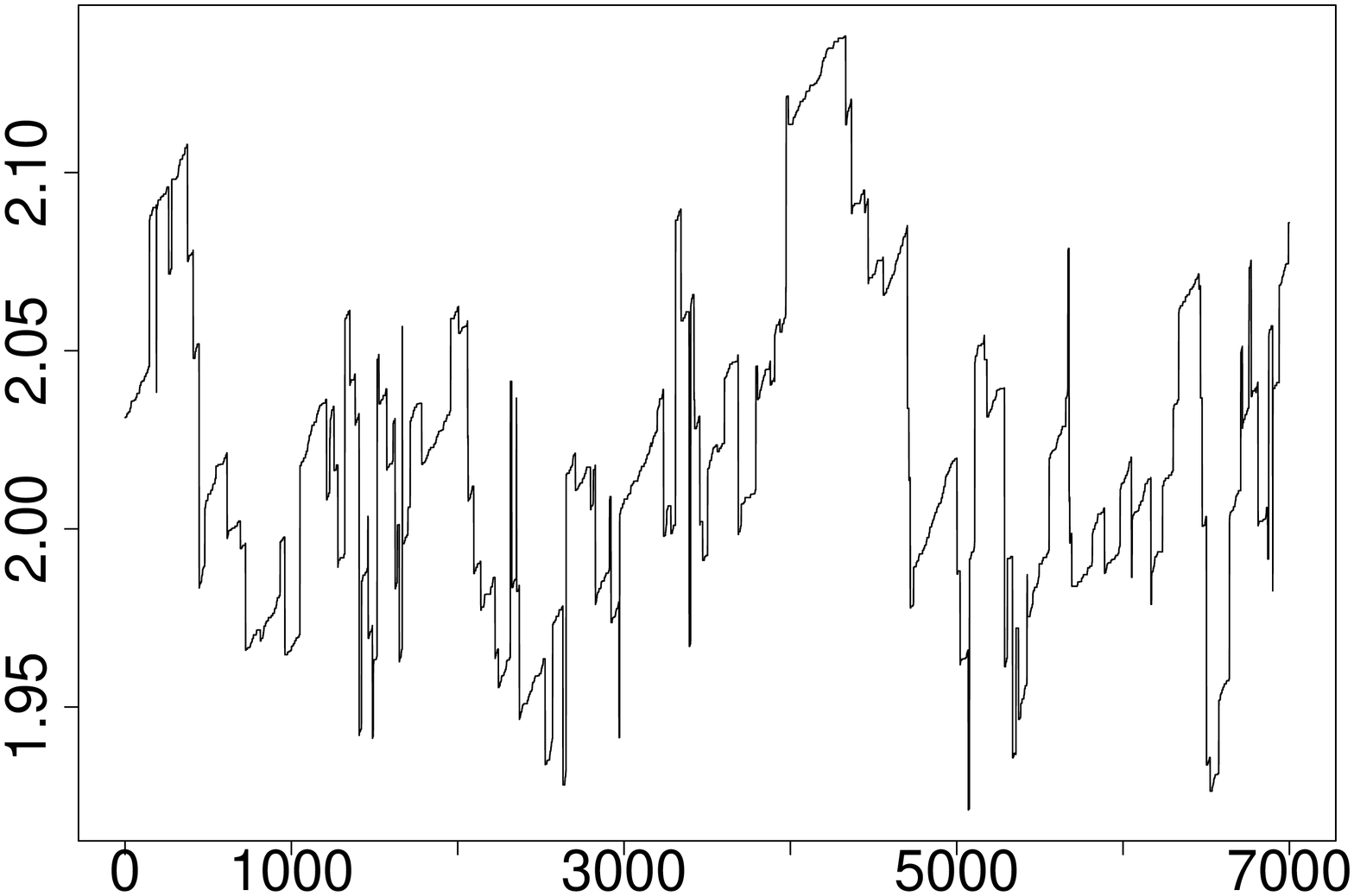}
      & \includegraphics[height = 4.5cm,angle=270]{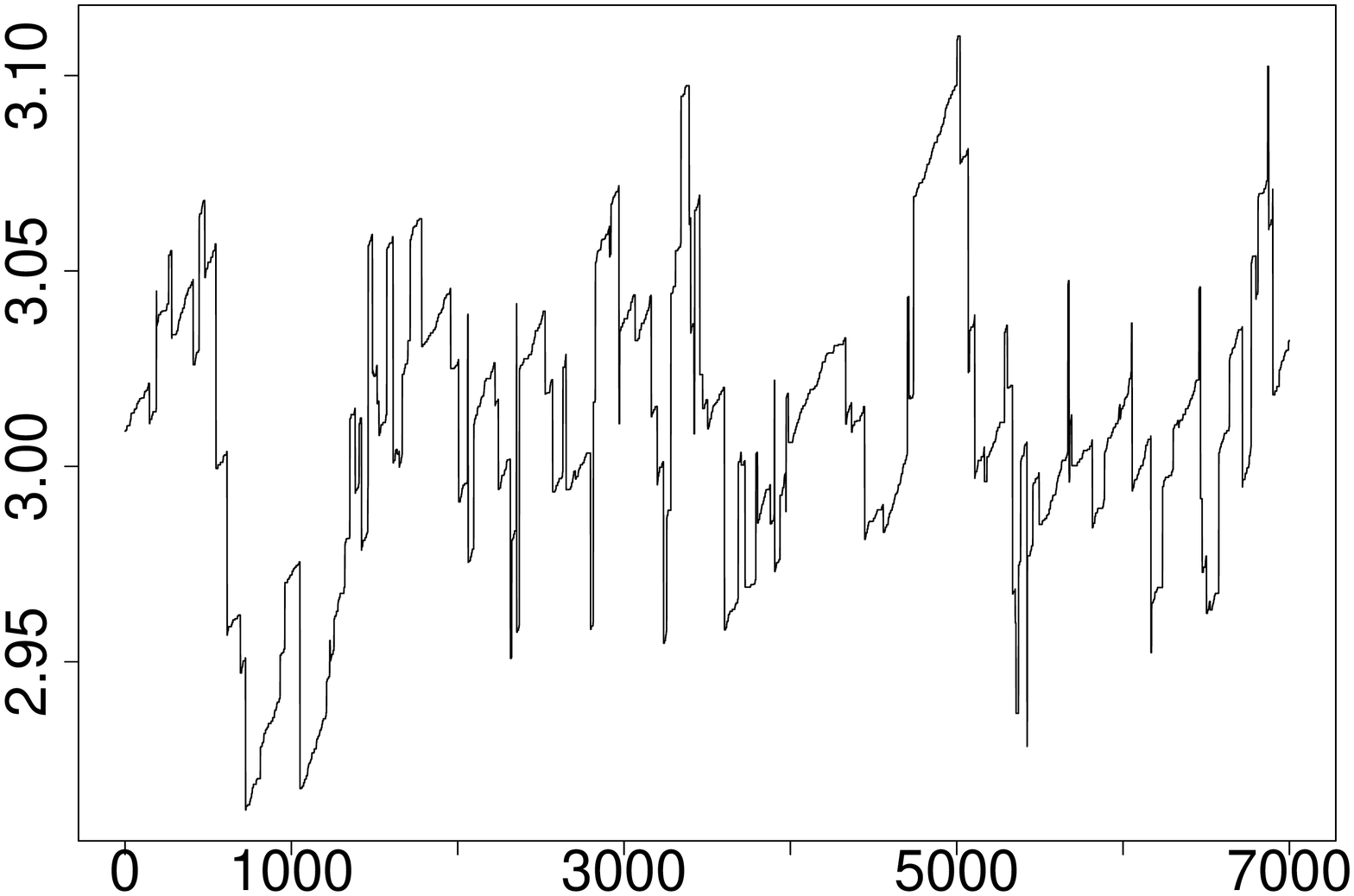} & \includegraphics[height = 4.5cm,angle=270]{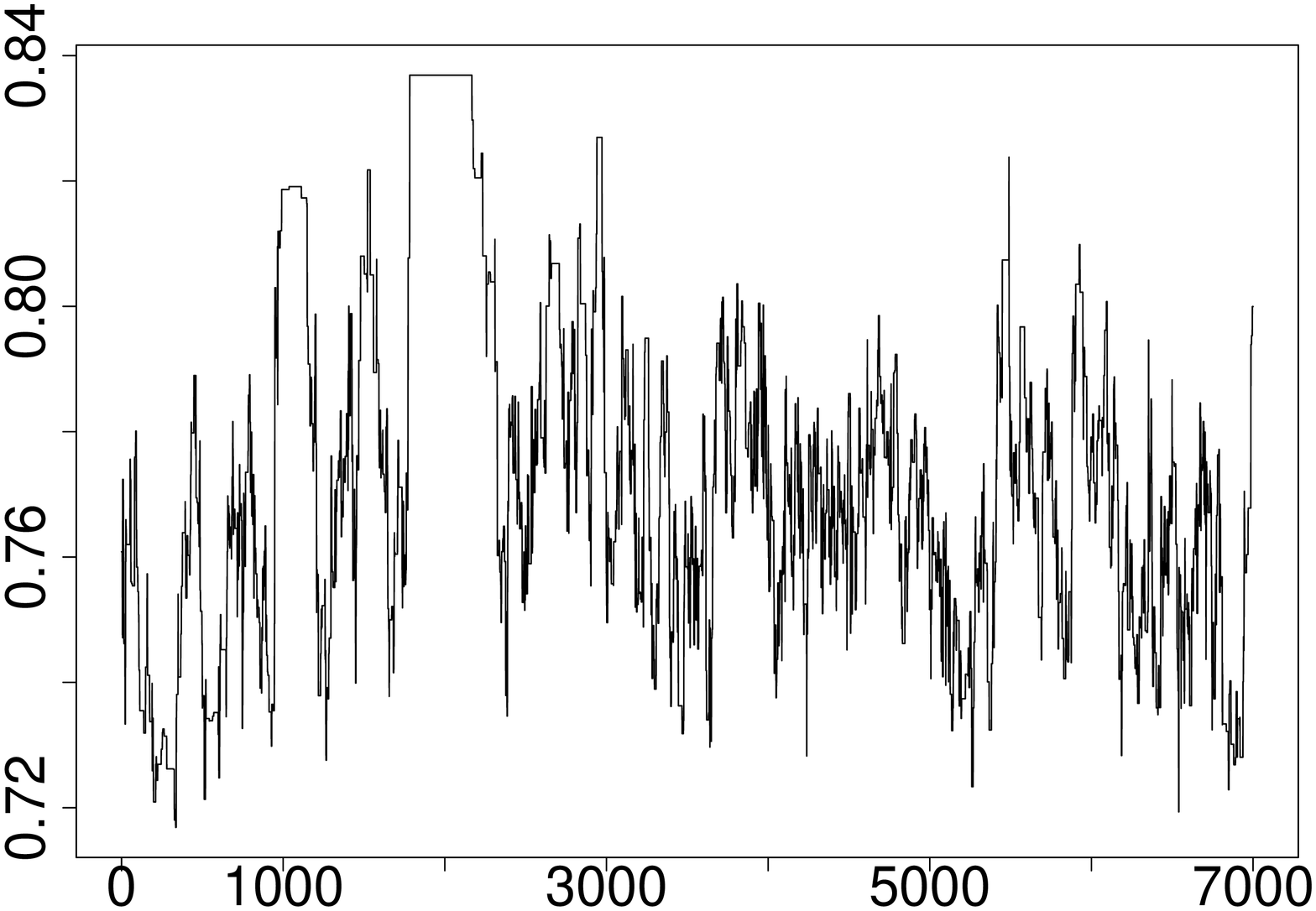} \\
		\hline
    Acf Plots & \includegraphics[height = 4.5cm,angle=270]{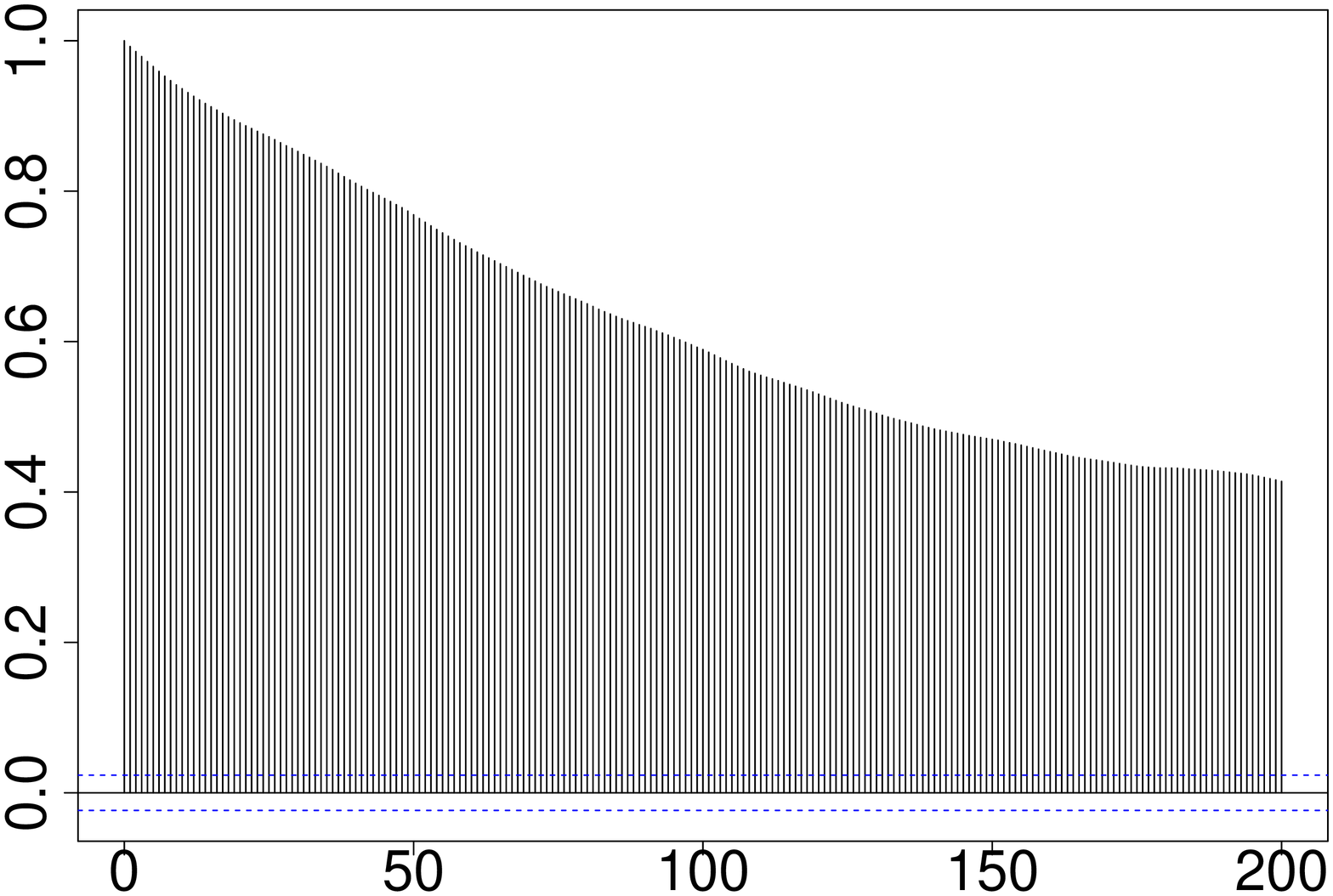}
      & \includegraphics[height = 4.5cm,angle=270]{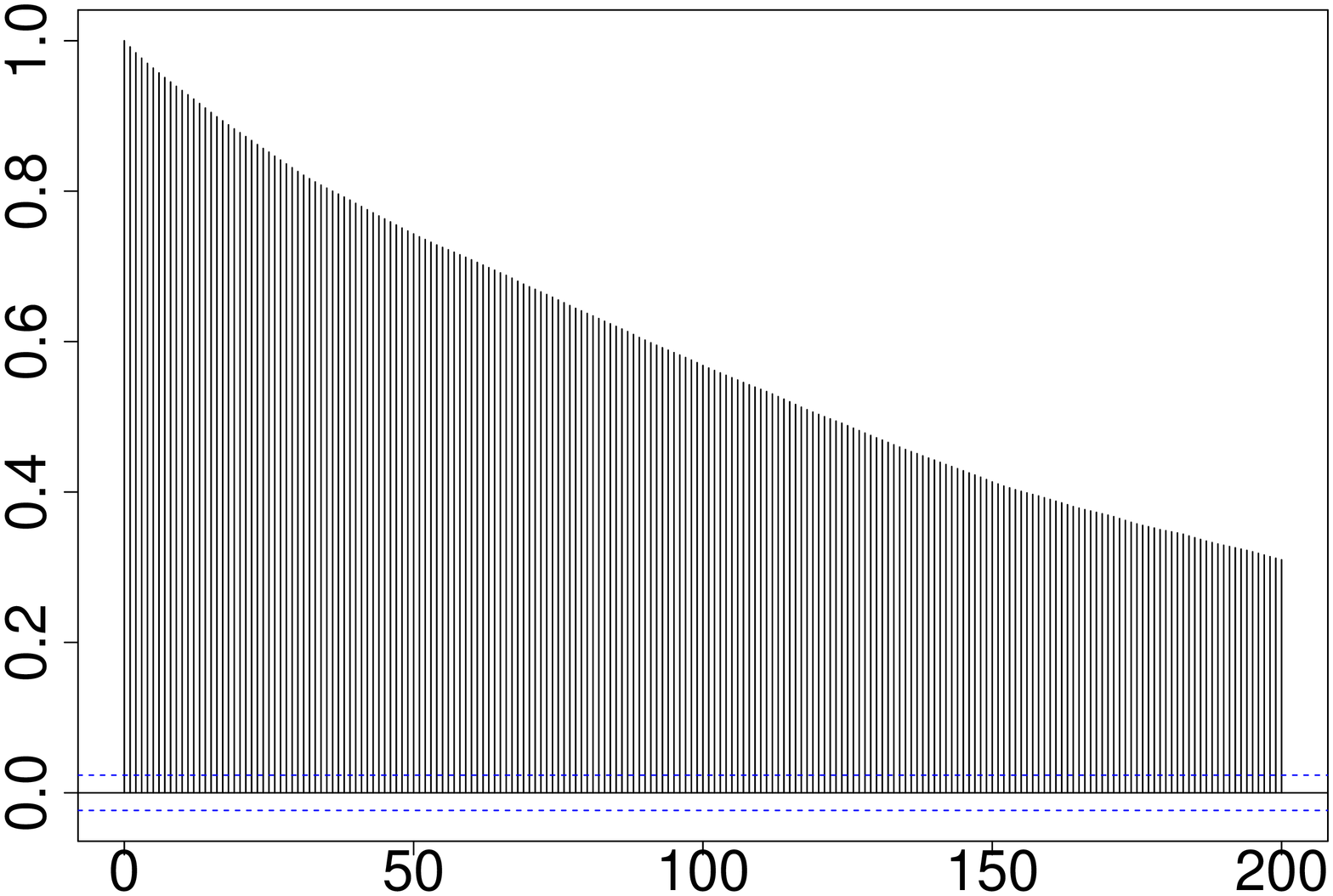} & \includegraphics[height = 4.5cm,angle=270]{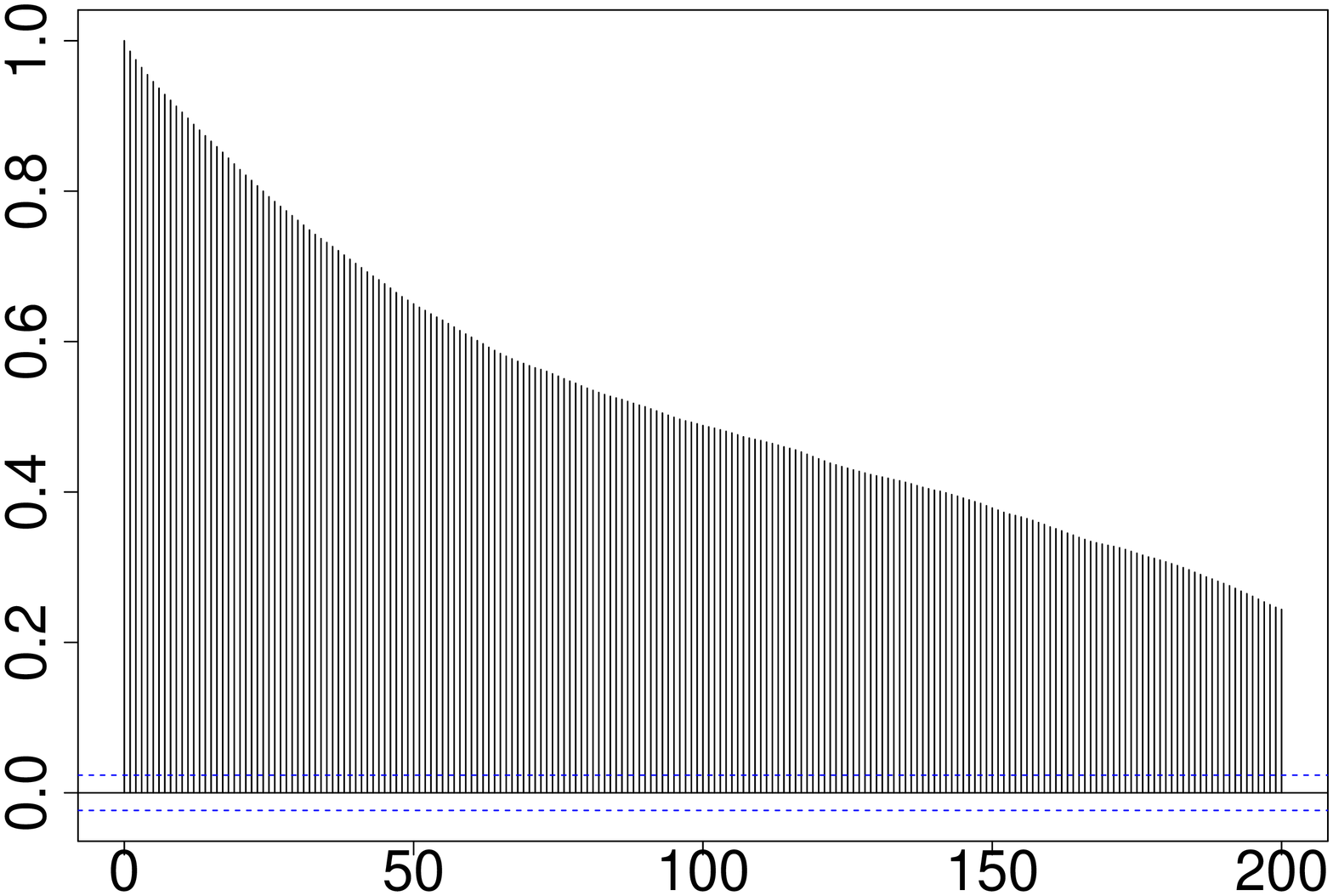} \\
		\hline
    Histogram & \includegraphics[height = 4.5cm,angle=270]{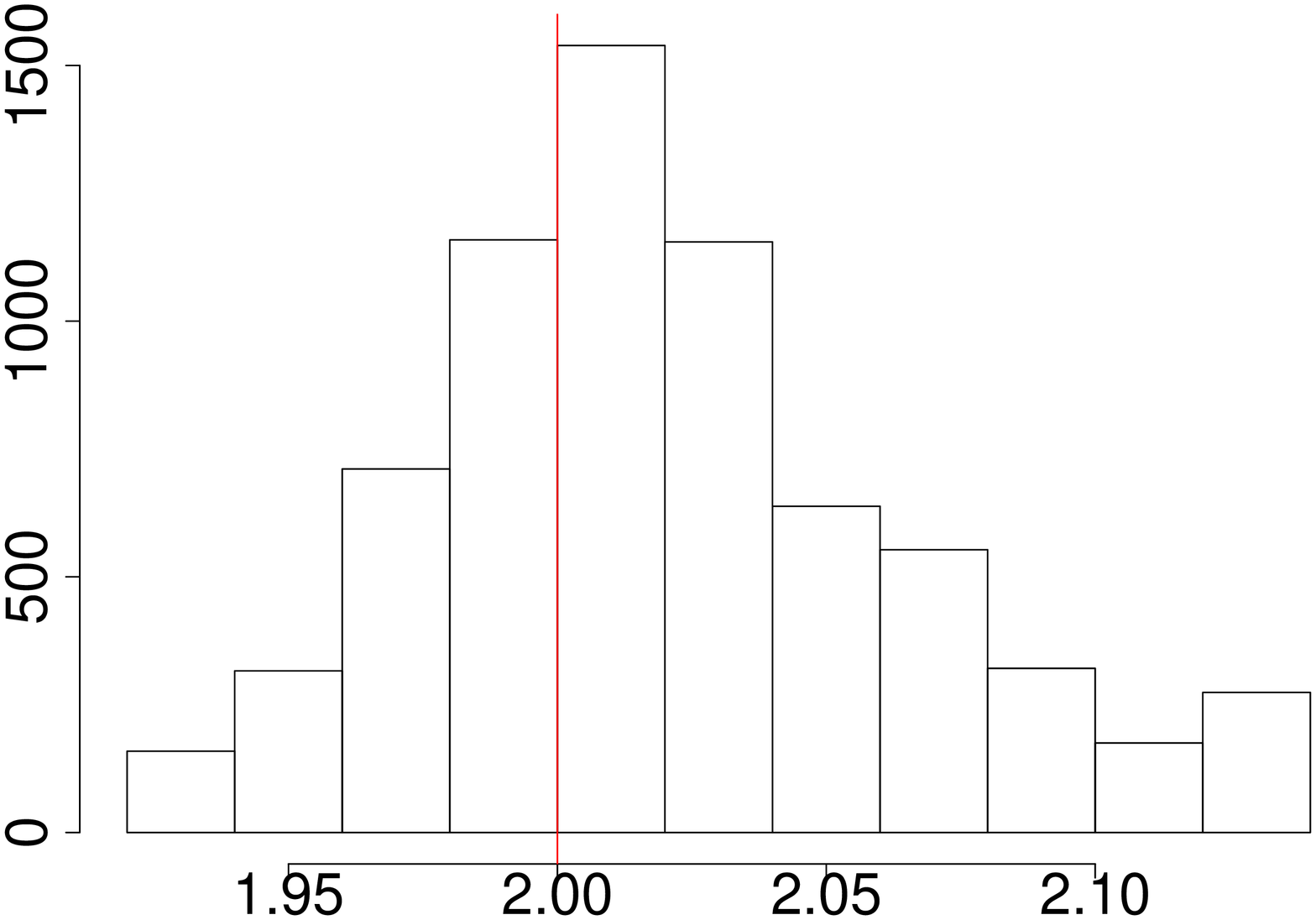}
      & \includegraphics[height = 4.5cm,angle=270]{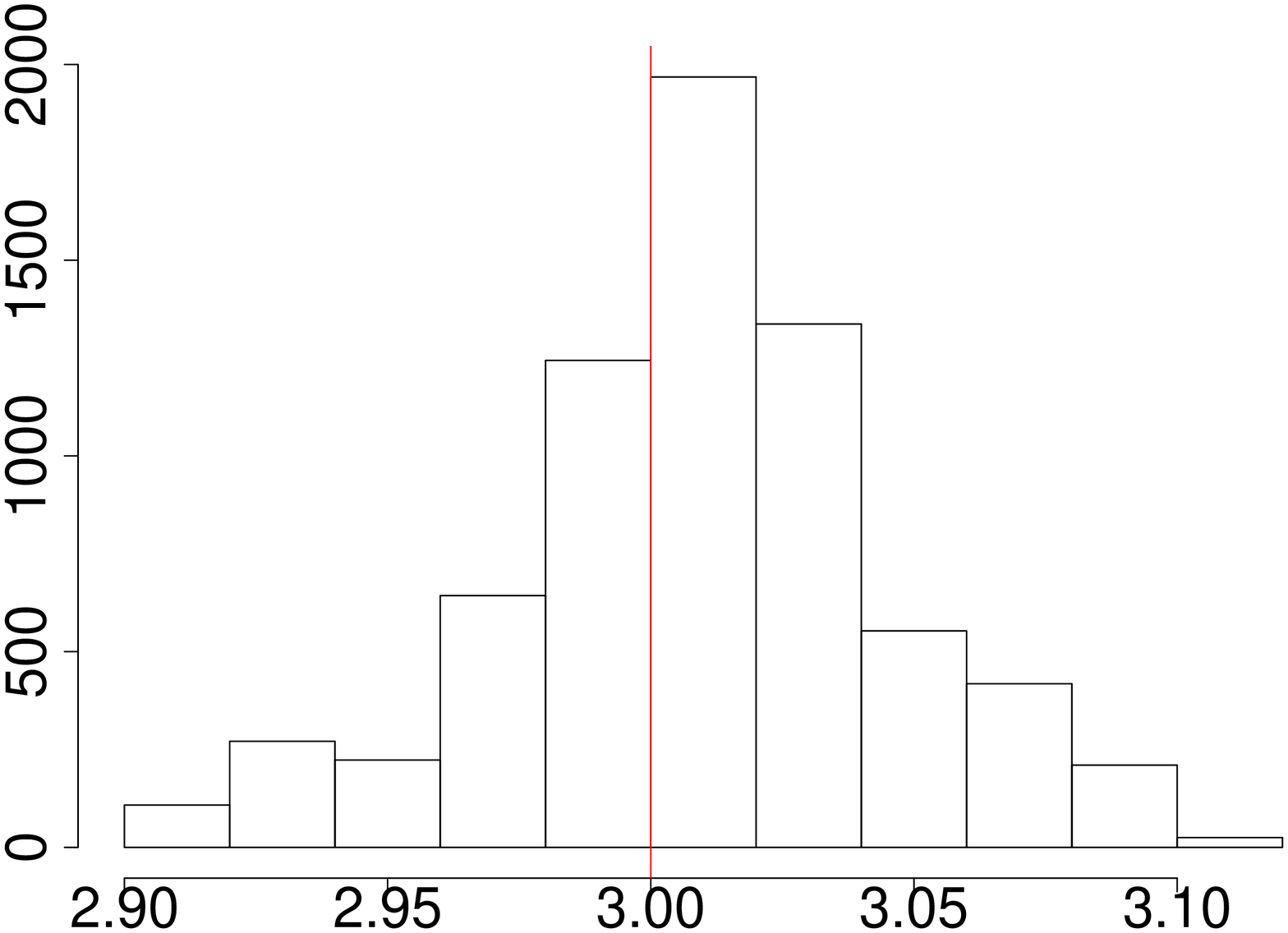} & \includegraphics[height = 4.5cm,angle=270]{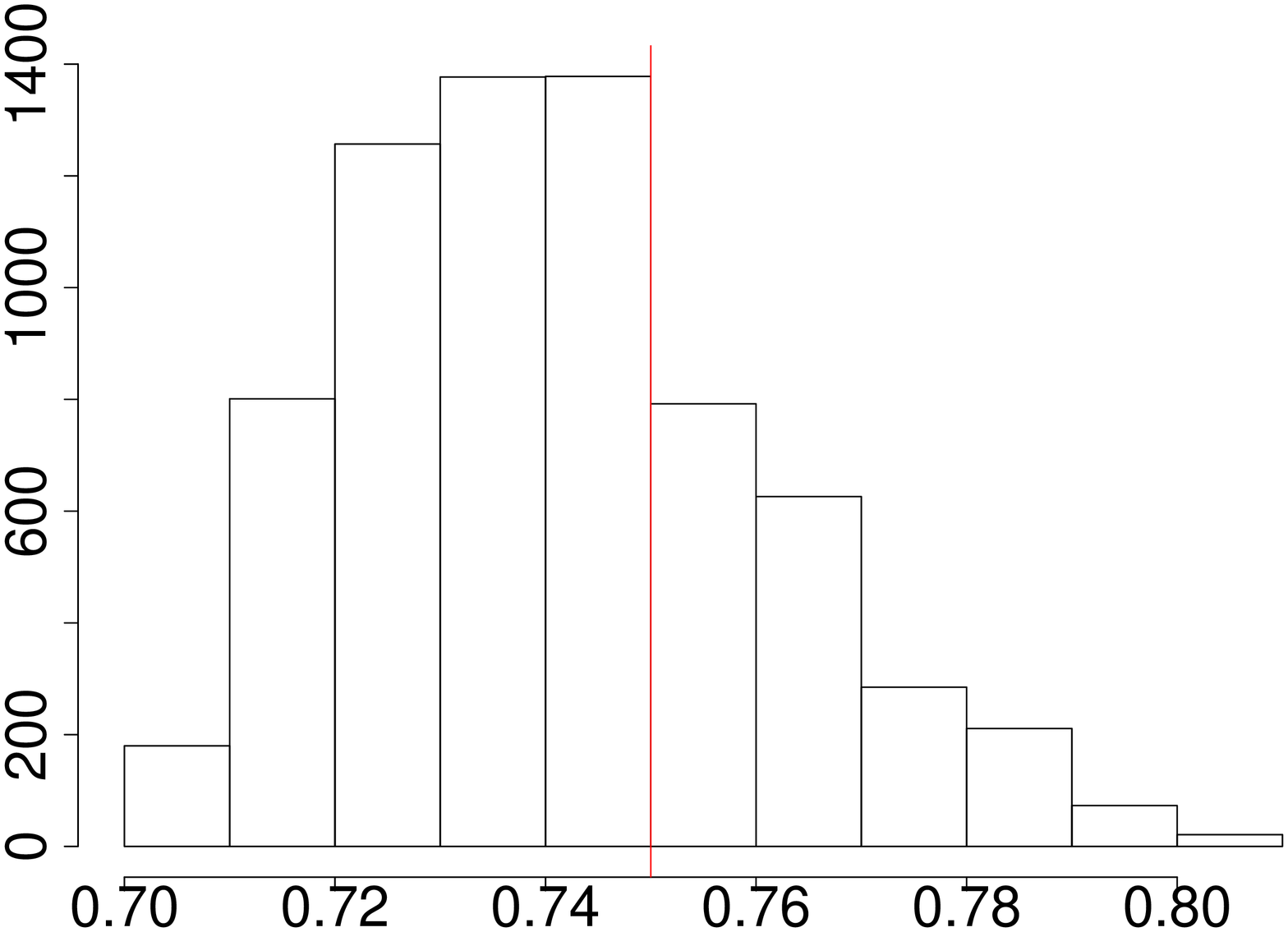} \\
		\hline
  \end{tabular}
  \caption{Trace (top row), ACF (middle row) and histogram (bottom row) plots for the regression coefficients of the first outcome, $\beta_{11}$ and $\beta_{12}$, and the corresponding residual standard deviation, $\sigma_1$.}\label{figtab2}
\label{fig:beta}
\end{table}

\begin{table}
\begin{center}
\begin{tabular}{|c|c|c|c|}
\hline
Criterion & Clayton & Frank & Gumbel\\
\hline
$\mbox{CVML}({\M_{2}})$ & -10213.4  & -7683.7 & -54763.2\\
$\mbox{CVML}({\M_{1}})$ & -7405.3 & -5569.4 & -49947.9\\ 
\hline
\end{tabular}
\caption{\it Twin Birth Data: CVML values for three copula families under model $\M_{1}$ (bottom row) and $\M_{2}$ (top row). The criterion suggests that the model $\M_{1}$ with the Frank copula is most suitable for fitting the data.  }
\label{twin-cvml}
\end{center}
\end{table}

\end{document}